\documentclass{IEEE-Con-Sys-mag}
\usepackage[utf8]{inputenc}
\usepackage{mathptmx} % assumes new font selection scheme installed

\let\labelindent\relax
\usepackage{enumitem}
\usepackage{dsfont}
\usepackage{url}
\usepackage{mathtools}

\usepackage{bbm}

\usepackage[compress]{cite}

\usepackage{tabularray}
\usepackage{multirow}
\usepackage{color, colortbl}

\usepackage[acronym]{glossaries}

\usepackage{graphicx}
\usepackage{adjustbox}
\usepackage[skip=5pt]{caption}
\usepackage{subcaption}
\usepackage{pgfplots}
\usetikzlibrary{plotmarks}
\usetikzlibrary{chains}
\usepackage{pgfplots}
\pgfplotsset{compat=newest}
\pgfplotsset{plot coordinates/math parser=false}
\newlength\figureheight
\newlength\figurewidth
\usetikzlibrary{shapes,arrows}
\usetikzlibrary{positioning}
\usetikzlibrary{decorations,decorations.pathmorphing,decorations.pathreplacing}
\usetikzlibrary{shadows,trees}
\pgfdeclaredecoration{switch}{initial}
{
    \state{initial}[width=+\pgfdecoratedinputsegmentremainingdistance]
    {
        \pgfsetlinewidth{1pt}
        \pgfpathcircle{\pgfpoint{0}{0}}{0.05\pgfdecoratedinputsegmentremainingdistance}
        \pgfpathlineto{\pgfpoint{0.25\pgfdecoratedinputsegmentremainingdistance}{0}}
        \pgfpathcircle{\pgfpoint{0.25\pgfdecoratedinputsegmentremainingdistance}{0}}%
            {0.05\pgfdecoratedinputsegmentremainingdistance}
        \pgfpathlineto{\pgfpoint{0.75\pgfdecoratedinputsegmentremainingdistance}{\pgfdecorationsegmentlength}}
        \pgfpathcircle{\pgfpoint{0.75\pgfdecoratedinputsegmentremainingdistance}{0}}%
            {0.05\pgfdecoratedinputsegmentremainingdistance}
        \pgfpathlineto{\pgfpoint{\pgfdecoratedinputsegmentremainingdistance}{0}}
        \pgfpathcircle{\pgfpoint{\pgfdecoratedinputsegmentremainingdistance}{0}}%
            {0.05\pgfdecoratedinputsegmentremainingdistance}
    }
    \state{final}
    {
        \pgfpathmoveto{\pgfpointdecoratedpathlast}
    }
}

\DeclareMathOperator*{\sbjto}{subject\ to}

\newcommand{\R}{\mathds{R}}
\newcommand{\Nz}{\mathds{N}_0}
\newcommand{\N}{\mathds{N}}
\newcommand{\EE}{\mathds{E}}
\newcommand{\PP}{\mathds{P}}
\newcommand{\inprod}[2]{\left\langle{#1}, {#2}\right\rangle}
\newcommand{\bmat}[1]{\begin{bmatrix}#1\end{bmatrix}}
\newcommand{\norm}[1]{\left|#1\right|}
\newcommand{\abs}[1]{\left|#1\right|}

\newcommand{\transp}{^\top}

\newcommand{\st}{x}

\newcommand{\meas}{y}

\newcommand{\control}{u}
\newcommand{\controlset}{\mathds{U}}
\newcommand{\wnoise}{w}
\newcommand{\mnoise}{\varsigma}

\newcommand{\Let}{\coloneqq}
\newcommand{\teL}{\eqqcolon}

\newcommand{\costps}{c_{\mathrm{s}}}
\newcommand{\costfinal}{c_{\mathrm{f}}}

\newacronym{mpc}{MPC}{Model Predictive Control}
\newacronym{smpc}{SMPC}{stochastic MPC}
\newacronym{tv}{TV}{time varying}
\newacronym{ti}{TI}{time invariant}
\newacronym{socp}{SOCP}{stochastic optimal control problem}
\newacronym{uq}{UQ}{uncertainty quantification}
\newacronym{pct}{PCT}{polynomial chaos theory}
\newacronym{pce}{PCE}{polynomial chaos expansion}
\newacronym{jcc}{JCC}{joint chance constraint}
\newacronym{gpc}{gPC}{generalized polynomial chaos}
\newacronym{lqr}{LQR}{linear quadratic regulator}
\newacronym{lmi}{LMI}{linear matrix inequality}
\newacronym{nn}{NN}{neural network}
\newacronym{pinn}{PINN}{physics informed neural network}
\newacronym{gp}{GP}{Gaussian process}
\newacronym{clot}{CLOT}{combined L-one and two}

\newtheorem{assumption}{Assumption}

\jvol{XX}
\jnum{XX}
\paper{8}
\jmonth{June}
\jname{IEEE CONTROL SYSTEMS}
\pubyear{2020}

\begin{document}

\title{Polynomial Chaos-based Stochastic Model Predictive Control\stitle{An Overview and \break Future Research Directions}}

\author{{P}rabhat K. Mishra, Joel A. Paulson, and Richard D. Braatz}
\affil{Indian Institute of Technology Kharagpur, University of Wisconsin-Madison, Massachusetts Institute of Technology }

\maketitle

\dois{}{}

\chapterinitial{M}odel Predictive Control (\acrshort{mpc}) has gained popularity due to its ability to handle constraints and incorporate model-based predictions \cite{2014_mpc_survey_Mayne}. While MPC exhibits inherent robustness through its receding horizon implementation and feedback mechanism \cite{2011_inherent_robustness_conditions, 2014_inherent_robustness}, this robustness is often limited in practical settings where uncertainties such as model mismatch, unmodeled dynamics, and fast-varying disturbances are significant. As highlighted by \cite{2004_nonrobustness_examples}, nominal stability guarantees do not automatically imply robustness under perturbations or constraint tightening, especially when feasibility and stability margins are narrow. Recent results on inherent robustness are applicable when states are within the robust positively invariant sets and assumptions on maximal increase in cost are satisfied in nominal MPC \cite[Assumptions 6 \& 7]{2021_reviewer_recommendation_8} and in economic MPC \cite[Assumption 6]{2023_reviewer_recommendation_14}. The inherent stochastic robustness \cite[Assumptions 3 \& 4]{2021_reviewer_recommendation_12} is also based on similar assumptions. It has been observed in \cite{2025_performance_nominal} that nominal MPC may lead to performance degradation or constraint violations when operating in uncertain or changing environments.

Therefore, robust MPC is developed by considering worst case bounds of uncertainties in the design stage and by ensuring desired properties for all possible realizations of uncertainties within those bounds \cite{2007_robust_mpc_survey}. However, the bounded description of uncertainties lead to a potentially very conservative MPC formulation in the presence of outliers or uncertainties with small chances of occurrences. As an alternative, Stochastic MPC (SMPC) overcomes conservatism by associating a probability distribution with the uncertainty \cite{2022_reviewer_recommendation_13}. 

\begin{summary}
	\summaryinitial{M}odern technology, from self-driving cars to the rapid production of life-saving vaccines, relies on the ability of machines to make intelligent decisions in a world full of uncertainty. Traditionally, engineers designed control systems using ``worst-case'' scenarios. While safe, this approach is often inefficient and limits the performance of high-tech devices. 
	In the past decade, polynomial chaos theory (PCT) has been shown to provide a computationally tractable way to perform complete and accurate uncertainty propagation through (smooth) nonlinear dynamic systems. As such, it represents a very useful computational tool for accelerating the computations needed in stochastic model predictive control (SMPC) with time invariant uncertainties. It turns out that it can also be used to reduce complexity of chance constraints, which are an important component of SMPC. In this paper, we provide an overview of PCT and discuss how it can be applied in SMPC.
\end{summary}

\begin{sidebar}{Uncertainty representation}
	\setcounter{sequation}{0}
	\renewcommand{\thesequation}{S\arabic{sequation}}
	\setcounter{stable}{0}
	\renewcommand{\thestable}{S\arabic{stable}}
	\setcounter{sfigure}{0}
	\renewcommand{\thesfigure}{S\arabic{sfigure}}
	\sdbarinitial{W}hile the importance of accounting for uncertainty during feedback control design has been long established, there is no consensus on the form in which the uncertainty should be represented. An expressive framework for modeling uncertainty can be a joint distribution of all uncertainties over time, constituting a non-Markovian stochastic process where the distribution of future uncertainties can depend on the complete or finite history of past uncertainty realizations. We often consider two special cases: (1) the uncorrelated process, where the autocorrelation function is a Dirac delta function, and (2) the fully correlated process, where the autocorrelation function is constant. Case (1) is typically known as the \emph{independent time-varying} (TV) case \cite{2022_reviewer_recommendation_10}, and case (2) is the \emph{dependent time-invariant} (TI) case \cite{2015_paulson_ti}. Most of the work on SMPC is concentrated on the TV case since it enables the use of the Markov property, greatly simplifying uncertainty propagation through the linear dynamical systems.  There are many practical problems in which the TV representation is conservative, making the TI representation a more realistic choice \cite{2015_paulson_ti}. For example, manufacturing variations \cite{2019_manufacturing_variations} lead to uncertain parameters, which remain static with time \cite{2023_ti} and their TI representation will be more realistic. We refer readers to \cite{2018_Constant_uncertainty} for more discussion on time-invariant uncertainties.
	
	The value of the parametric uncertainty, denoted as $\theta$, can vary across experiments depending on the specific application. This variability might arise, for instance, when new raw materials are introduced into a manufacturing system. This phenomenon has been extensively discussed in various literature sources in \cite{PCE_NN_2022, review_IC_PCE_2018}. For example, in an armature-controlled direct-current electric motor with independent excitation \cite{2018_robustness_analysis}, the moment of inertia and mechanical friction are the functions of the load, the armature resistance is affected by temperature variations, and the motor constant is a function of the field magnetic flow. 
	In some applications, the assumption that $\theta$ is time-invariant is only an approximation, and in reality the value of $\theta$ is time-varying but over a time scale that is longer than the time scales of the closed-loop dynamics. In such cases, designing a controller that is robust to the parametric uncertainty can be practically useful. We refer readers to \cite{1987_comparison_robust_adaptive, 2001_limitation_adaptive_robust} for comparative studies between robust and adaptive controllers.
\end{sidebar}

SMPC has emerged as a way to systematically incorporate a probabilistic description of uncertainties in the framework of stochastic optimal control problems (SOCPs) involving cost and constraints \cite{2017_Rawlings_book,2016_review_SMPC}. The fundamental tools of SMPC come from stochastic programming and chance-constrained optimization.
It is illustrated in \cite{2022_reviewer_recommendation_11} that certain SMPC formulations satisfying interesting properties such as exponential stability in expectation \cite{2021_reviewer_recommendation_9} and distributional robustness \cite{2023_reviewer_recommendation_15} may not necessarily be more robust than nominal MPC in the presence of bounded disturbances.

Although SMPC is a conceptually simple and intuitive extension of robust MPC, it introduces a couple of important challenges: (i) the representation of the uncertainties and (ii) their efficient propagation through the dynamics. The uncertainty in the output $y=f(x)$ due to the uncertainty in $x$ is called uncertainty propagation through $f$.

\begin{figure}[b]
	\centerline{\includegraphics[width=18.0pc]{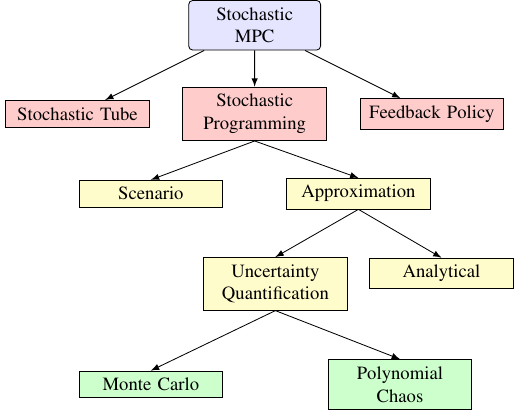}}
	\caption{Classification of stochastic MPC}
	\label{MPC_classification}
\end{figure}

The literature on stochastic MPC can be classified according to the basic theme of approaches in Fig. \ref{MPC_classification} and an overview is available in \cite{2016_review_SMPC}. The stochastic tube approaches \cite{stochastic_tube_2010} are proposed as generalization of deterministic tube MPC \cite{2005_linear_tube} by considering the probabilistic description of uncertainties. On the other hand, feedback policy-based approaches \cite{policy_2008, hokayem2009stochastic, 2011_TAC_Lygeros} are mainly focused on providing computationally tractable feedback policies and can be combined with other approaches of stochastic MPC. The stochastic programming-based SMPC can be broadly divided into the \textit{scenario approach}, which aims to construct randomized solutions with formal guarantees on constraint satisfaction, and \textit{approximation-based methods}, which rely on analytical or uncertainty quantification (UQ)-based approximations of the cost and constraint functions.
In the scenario approach to SMPC \cite{prandini2012randomized}, a finite number of disturbance realizations are sampled and used to enforce deterministic constraints that replace the original chance constraints \cite{scenario_2006, scenario_2009, calafiore_2012, 2009_scenario_Bemporad}. Unlike standard sampling-based methods, the scenario approach incorporates a structured procedure: after solving the problem with all sampled constraints, a subset may be discarded to improve performance, provided that the remaining empirical violation level stays within a user-defined tolerance. This process yields a solution that (with high probability) satisfies the original chance constraints at the desired violation level. Importantly, the required number of samples can be determined \textit{a priori} based on theoretical sample complexity bounds, and the approach remains valid regardless of the underlying uncertainty distribution.

A conceptually simpler alternative is the use of Monte Carlo simulation to approximate expectations or chance constraints directly \cite{batina_2004_monte-carlo}. Here, the control problem is solved using a fixed set of randomly sampled disturbances, with no post-processing or guarantees on feasibility. While both the scenario and Monte Carlo approaches rely on random sampling, the latter lacks the built-in certification mechanisms of the former and treats the sample size as a heuristic design choice.
Beyond sampling-based methods, approximation-based SMPC also includes approaches that seek closed-form or tractable approximations of the stochastic objectives and constraints. These include analytical reformulations \cite{hewing_2018} as well as UQ techniques such as polynomial chaos theory (PCT). PCT approximates stochastic variables using orthogonal polynomial expansions, enabling efficient propagation of uncertainty and direct access to statistical moments (as well as the construction of ordinary differential equations that directly capture the time-evolution of the moments in continuous-time systems \cite{2017_Mesbah_IJRNC}).

PCT was originally developed by Norbert Wiener \cite{wiener_1938_homogeneous} as a way to construct a stochastic approximation of the mathematical model that quantifies the effect of the uncertainties on states and outputs without using online sampling. Decades later, researchers began to incorporate PCT into the formulation of open- and closed-loop optimal control problems. The PCT-based approach is generally applicable and has long been applied to both static and dynamic models to carry out quantitative risk analysis and ensure a safety margin in the operation or design of civil, environmental,  mechanical, and electronic systems including power systems \cite{power_system_2004, 2016_power_system}, integrated circuits \cite{review_IC_PCE_2018}, fluid dynamics \cite{review_UQ_Fluid_2009}, aerospace \cite{Satellite_2013}, dynamic flux balance analysis \cite{2019_nonsmooth_PCT} and continuous lyophilization process \cite{2025_braatz_pc}. A tutorial introduction of PCT and a high-level review of the PCT-based control results up to 2012 is available \cite{Kim_magazine_2013}.

\begin{pullquote}
	A PCE is an approximation of the implicit function between input $\theta$ and output $y$ with an explicit function, a linear combination of \emph{polynomial basis functions}.
\end{pullquote}

This article, in contrast, is devoted to providing a review of mathematical formulations in which PCT has been incorporated into stochastic model predictive control. In the past decade, PCT has been shown to provide a computationally tractable way to perform complete and accurate uncertainty propagation through (smooth) nonlinear dynamic systems. As such, it represents a very useful computational tool for accelerating the computations needed in SMPC with TI uncertainties. It turns out that it can also be used to reduce complexity of chance constraints, which are an important component of SMPC. In this paper, we provide an overview of PCT and discuss how it can be applied in such TI settings. 

This review is timely for several reasons. First, many contributions have been made to PCT-based MPC in recent years, which are not covered in past reviews \cite{Kim_magazine_2013, 2016_review_SMPC}. Second,  PCT-based MPC is increasingly applied in control applications including
autonomous driving \cite{Autonomous_navigation_Jones_2021}, motion planning \cite{motion_planning_2023}, haptic collaboration \cite{haptic_20}, Li-ion battery \cite{EV_battery_22}, 
propulsion system \cite{propulsion_system_22}, petroleum reservoir \cite{Reservoir_22}, fault diagnosis \cite{mesbah_2014_diagnosis}, and continuous pharmaceutical manufacturing \cite{2018_pharmaceutical}. Our notations are standard. All acronyms are defined before their first use and they are also summarized in the Table \ref{tab:acronym}.

\begin{table}[h]
  \centering
  \caption{\label{tab:acronym}Acronyms}
  \begin{tblr}{
    colspec = {|c|c|},
    row{1} = {blue!15},
  }
    \hline
      Acronym &  Phrase \\
    \hline
\acrshort{mpc}& \acrlong{mpc}\\ \hline
\acrshort{smpc} & \acrlong{smpc}\\ \hline
\acrshort{tv} & \acrlong{tv}\\ \hline
\acrshort{ti} & \acrlong{ti}\\ \hline
\acrshort{socp} & \acrlong{socp}\\ \hline
\acrshort{uq} & \acrlong{uq}\\ \hline
\acrshort{pct} & \acrlong{pct}\\ \hline
\acrshort{pce} & \acrlong{pce}\\ \hline
\acrshort{jcc} & \acrlong{jcc}\\ \hline
\acrshort{gpc} & \acrlong{gpc}\\ \hline
\acrshort{lqr} & \acrlong{lqr}\\ \hline
\acrshort{lmi} & \acrlong{lmi}\\ \hline
\acrshort{nn} & \acrlong{nn}\\ \hline
\acrshort{pinn} & \acrlong{pinn}\\ \hline
\acrshort{gp} & \acrlong{gp}\\ \hline
\acrshort{clot} & \acrlong{clot}\\ 
\hline   
  \end{tblr}
\end{table}

\section{Stochastic Model Predictive Control (SMPC): The General Case}\label{s:MPC}
We consider a stochastic discrete-time uncertain nonlinear system described by
\begin{equation}\label{e:general_system}
	\begin{aligned}
		\st_{t+1} &= f(\st_t, \control_t, \wnoise_t, \theta), \\
		\meas_t &= h(\st_t, \mnoise_t, \theta), 
	\end{aligned}
\end{equation}
where $\st_t \in \R^{n_x}$, $\control_t \in \controlset \subset \R^{n_\control}$, and $\meas_t \in \R^{n_\meas}$ are the states with initial value $\st_0$, actions, and measured outputs of the system at time $t\in\Nz$, respectively. The process noise $\wnoise_t$, the measurement noise $\mnoise_t$, and parametric uncertainty $\theta$ have appropriate dimensions and known distributions $\rho_\wnoise$, $\rho_\mnoise$, and $\rho_\theta$, respectively. The initial state $x_0$ also has a known density $\rho_{x_0}$.
Further, we make the relatively mild assumption \cite[Assumption 1 \& 2]{2018_Automatica_Bitmead} that the one-step transition function $f$ and measurement function $h$ are both Borel measurable. 
\begin{assumption}[\cite{2018_Automatica_Bitmead}]\label{as:mdp}
	The dynamics \eqref{e:general_system} satisfy
	\begin{enumerate}
		\item $f(\cdot, \control, \cdot, \cdot)$ is differentiable almost everywhere with full rank Jacobian for all $\control \in \R^{n_u}$.
		\item $h(\cdot, \cdot, \cdot)$ is differentiable almost everywhere with full rank Jacobian.
		\item $x_0, \wnoise_i, \mnoise_j, \theta$ are mutually independent and identically distributed for all $i,j \geq 0$.
		\item The control input $u_t$ at time instant $t\geq 0$ causally depends on the data $D_t \Let \{\meas_t, \control_{t-1}, \meas_{t-1} \ldots, \control_0, \meas_0 \}$, and prior density $\rho_{x_0} \Let \text{pdf}(\st_0 \mid D_{-1})$.
	\end{enumerate}
\end{assumption}
The above assumption ensures that the system \eqref{e:general_system} is a controlled Markov process \cite{kumar1986stochastic} in the absence of parametric uncertainty $\theta$ and admits the analysis of stochastic behavior through sampling based methods. They can be considered as regularity conditions allowing the development of Bayesian filter \cite{2018_Automatica_Bitmead}.

\begin{figure}[b]
	\centerline{
		\includegraphics[width=18.0pc]{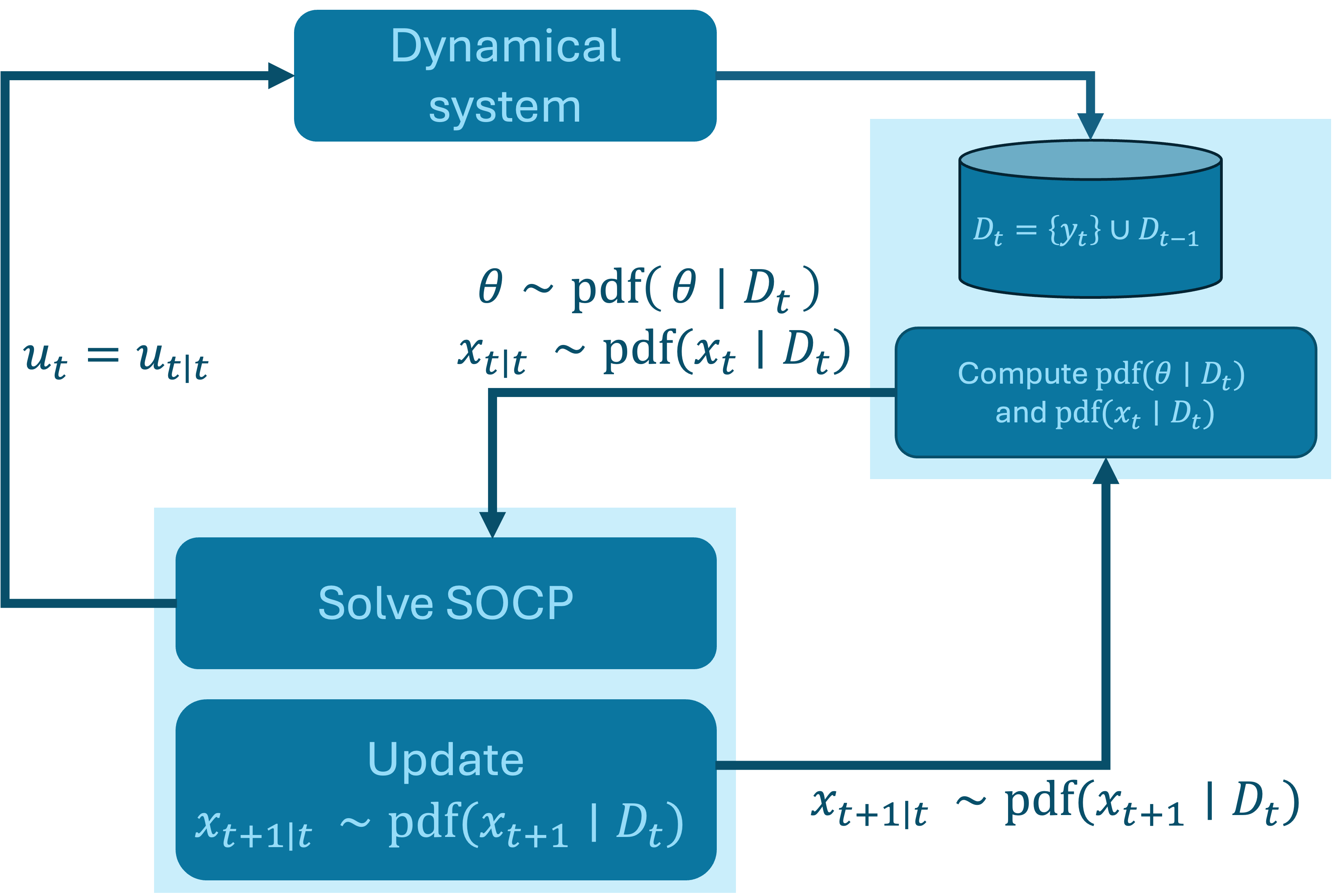}}
	\caption{General SMPC algorithm}
	\label{fig:smpc_general}
\end{figure}

The conditional density of the state $\st_t$ given data $D_t$,
$\text{pdf}(\st_t \mid D_t)$
is called the information state of the system \eqref{e:general_system}, which is propagated by
\begin{equation}\label{e:Bayesian_filter}
	\begin{aligned}
		\text{pdf}(\st_t \mid D_t) &= \frac{\text{pdf}(\meas_t \mid \st_t) \text{pdf}(\st_t \mid D_{t-1})}{\int \text{pdf}(\meas_t \mid \st_t) \text{pdf}(\st_t \mid D_{t-1})d \st_t},  \\
		\text{pdf}(\st_{t+1} \mid D_t) &= \int \text{pdf}(\st_{t+1} \mid \st_t, \control_t) \text{pdf}(\st_t \mid D_t) d \st_t, 
	\end{aligned}
\end{equation}
for $t \in \Nz$ and prior density $\rho_{\st_0}$ as in Assumption \ref{as:mdp}. The recursion \eqref{e:Bayesian_filter} is called the Bayesian filter, which is equivalent to Kalman filter for the Linear systems with Gaussian noise. Similarly, Bayes' rule can give the posterior $\text{pdf}(\theta \mid D_t)$ in terms of the likelihood $\text{pdf}(\meas_t \mid \theta)$, prior $\text{pdf}(\theta \mid D_{t-1})$ and evidence (or marginal) $\text{pdf}(\meas_t \mid D_{t-1})$ by
\begin{equation}\label{e:baye's_rule}
	\text{pdf}(\theta \mid D_t) = \frac{\text{pdf}(\meas_t \mid \theta)\text{pdf}(\theta \mid D_{t-1})}{\text{pdf}(\meas_t \mid D_{t-1})}.
\end{equation}

We first fix an optimization horizon $N \in \N$ and define a feedback policy $\pi_i: \R^{(i+1)n_x} \rightarrow \controlset \subset \R^{n_u}$ for $i=0, \ldots, N-1$. Let us define the cost-per-stage function $\costps:\R^{n_x}\times\controlset \rightarrow \R_{\geq 0} $ and the final cost function \(\costfinal:\R^{n_x} \rightarrow \R_{\geq 0} \). The objective function for the stochastic optimal control is 
\begin{equation}\label{e:expected_cost}
	V(\st_{t\mid t}, \pi) \Let \EE \left[ \costfinal(\st_{t+N\mid t}) + \sum_{i=0}^{N-1} \costps(\st_{t+i\mid t}, \control_{t+i\mid t})  \right],  
\end{equation}
where $\st_{t \mid t} \sim \text{pdf}(\st_{t}  \mid D_t ), \control_{t+i\mid t} = \pi_i(\st_{t\mid t}, \cdots, \st_{t+i\mid t})$, $\pi = [\pi_0\transp \ \cdots \ \pi_{N-1}\transp]\transp \in \R^{n_\control N}$, and $z_{k \mid t}$ denotes the predicted value of the variable $z_k$ for $k\geq t$ predicted at time $t$. The expectation in \eqref{e:expected_cost} is taken over joint-distribution of uncertainties present in \eqref{e:general_system}. The choice of $\costps$ and $\costfinal$ allow penalization of higher order moments such as variance, skewness and kurtosis. For the practical applications of interest, hard constraints are considered on the control and soft constraints on states (or outputs) of the form
\begin{align}
	& \control_{t} \in \controlset \label{e:constraint_input} \\
	& \PP [ \st_t \in \mathcal{X} ]  \geq 1-\beta \label{e:jcc},
\end{align}
where $\beta \in  [0,1[$ represents the constraint violation probability and the constraint \eqref{e:jcc} is called the joint chance constraint (JCC). The control set $\controlset$ defines the upper and lower bounds of the actuators. The set $\mathcal{X}$ is assumed to be a polytope,
\begin{equation}\label{e:polytopic_state_constraint}
	\mathcal{X} \Let \{\st \in \R^{n_\st} \mid a_\ell^\top \st \leq b_\ell  \text{ for } \ell = 1, \ldots, n_c  \}.
\end{equation}
In this case, $n_c$ is number of inequality constraints probabilistically and jointly satisfied with the probability of at least $1-\beta$. The centerpiece of SMPC is the constrained SOCP:
\begin{subequations}\label{e:general_SOCP}
	\begin{align}
		\min_{\pi} & \quad V(\st_{t\mid t}, \pi)  \label{e:objective}\\
		\sbjto & \quad \st_{t\mid t} \sim \text{pdf}( \st_t \mid D_t) \label{e:initialization}\\
		& \quad \st_{t+i \mid t} = f(\st_{t+i -1 \mid t}, \control_{t+i-1\mid t}, \wnoise_{t+i-1}, \theta_{t+i-1}); \notag \\ & \quad \text{ for } i = 1, \ldots, N \label{e:propagation}\\
		& \quad \control_{t+i \mid t} \in \controlset; \quad i = 0, \ldots, N-1 \\
		& \quad \PP [\st_{t+i\mid t} \in \mathcal{X}] \geq 1-\beta;\quad  i = 1, \ldots, N \label{e:socp_chance_constraint}.
	\end{align}
\end{subequations}

The SOCP \eqref{e:general_SOCP} requires the state estimation through \eqref{e:Bayesian_filter}. Readers may refer to \cite{2013_separation_principle} for the validity of separation principle in such problems, which are not discussed in this manuscript. A general SMPC algorithm without implementation details is given in Fig. \ref{fig:smpc_general}.

\begin{sidebar}{Key considerations in SMPC}
	\setcounter{sequation}{0}
	\renewcommand{\thesequation}{S\arabic{sequation}}
	\setcounter{stable}{0}
	\renewcommand{\thestable}{S\arabic{stable}}
	\setcounter{sfigure}{0}
	\renewcommand{\thesfigure}{S\arabic{sfigure}}
\sdbarinitial{T}he key considerations when solving the SOCP are:
	\begin{enumerate}[leftmargin = *]
		\item \textbf{Computational complexity associated with the uncertainty propagation through the dynamics \eqref{e:propagation}:} The solution of the SOCP \eqref{e:general_SOCP} involves prediction of states through \eqref{e:propagation}. The most popular methods of uncertainty propagation are based on linear approximation of the system dynamics \cite{2000_linearized_propagation} (typically relying on existence and calculation of Jacobians along all system trajectories), unscented transform \cite{2004_unscented_filtering} (propagates sigma points through the dynamics), PCT, Monte Carlo \cite{2006_mc, 2009_sequential_mc}, and the scenario approach \cite{2013_scenario, 2014_scenario}. The first three methods of uncertainty propagation are compared in \cite{uncertainty_propagation_comparison_2016} for a normal and a uniform distribution of parametric uncertainties. The authors highlighted the advantage of using PCT in terms of its ability to incorporate \emph{a priori} knowledge about the distribution. Another popular method to compute statistical properties is the quasi-Monte Carlo method. In a standard Monte Carlo,  $N$ samples of the uncertainty $\theta$ are drawn from its distribution $\rho_\theta$ and the output $y_i$ calculated for each $\theta_i$. Then empirical statistics are computed from the results, such as the expected value, $\EE [y] = \frac{1}{N} \sum_{i=1}^N y_i$, and the variance. This method requires many model evaluations, on the order of $10^3$--$10^4$ \cite{2014_PCE_Sudret}, to reach an acceptable accuracy. In a (quasi) Monte Carlo method, the number of model evaluations is reduced by selecting the samples through a low discrepancy sequence such as the Hammersley sequence. An advantage of PCT over Monte Carlo methods and the scenario approach is its computational efficiency, as reviewed later in this article.
		\item \textbf{Non-convexity of the general chance constraints \eqref{e:jcc}:} Even if the pdf of the states is available, the chance constraints \eqref{e:socp_chance_constraint} are difficult to implement because they are non-convex in general. For linear systems, the JCC can be reformulated analytically (when there exists only additive Gaussian noise) \cite{2012_oldewurtel}, via distributionally robust constraints \cite{2009_robust_optimization}, or via stochastic tubes (using backoff parameters offline) \cite{2015_MPC_book, 2016_constraint_tightening}. For nonlinear systems, the JCC has been approximated by discretization of probability distribution \cite{2000_discretized_pdf}, convex approximation \cite{2007_convex_approximation}, sample average approximation \cite{2009_saa, blackmore_2010_particle_chance}, and moment-based approximation \cite{2017_Mesbah_IJRNC}. JCCs are discussed in the context of PCT in this article.
		\item \textbf{Optimality of the SOCP \eqref{e:general_SOCP} may not imply the stability of the system \eqref{e:general_system}:} In robust MPC formulations, robust stability is guaranteed by ensuring stability for all possible realizations of uncertainties, such as by constructing robustly positively invariant sets. 
		Since the construction of a robust positively invariant set is not possible in the presence of uncertainties or disturbances with unbounded support, robust MPC methods to ensure stability are not directly applicable in stochastic MPC.
		Existing conditions of stability are reviewed in \cite{2018_stabilizing_conditions} along with limitations and challenges associated with stochastic stability \cite{2014_chatterjee_lygeros}. 
		\item \textbf{Ensuring recursive feasibility of \eqref{e:general_SOCP} in the presence of hard constraints on inputs \eqref{e:constraint_input}:} Since all realizations of uncertainties cannot be included in the stability analysis of a closed-loop system, at least for uncertainties with unbounded support, a plausible approach is to include stabilizing conditions as an additional constraint. Even if feasibility of such a constraint is assumed for the initial state, ensuring feasibility for all future states is very complicated, and complications increase  when the control authority itself is bounded as in practical systems. Given feasibility of the underlying optimization at the initial state, the property of its feasibility for all time is known as \emph{recursive feasibility} \cite{2009_primbs_recusrsive_feasibility}. Feasibility has been addressed by relaxing the constraints by following the approach of \cite{2006_chance_constraint_approximation} and in \cite{2017_mesbah} by optimal risk allocation, minimizing the constraint violation probability \cite{2021_constraint_violation}, adjusting the probability threshold \cite{2021_recursive_feasibility}, controlling the average constraint violation \cite{2014_korda_average_constraint_violation} and by discounting the constraints 
		\cite{policy_2008,2021_discounted_constraint}. A method has also been proposed \cite{kohler_2022} that replaces the constraint \eqref{e:initialization} by the interpolated constraint $\st_{t\mid t} = \lambda \st_t + (1-\lambda)\st_{t\mid t-1} $, where $\lambda$ is an interpolation variable.     
		\item \textbf{Tractability of the general form of the feedback policies $\pi$:} The optimization over feedback policies for stochastic systems has generally higher performance than open-loop control sequences \cite[\S 2.4]{kumar1986stochastic} but the optimization over all admissible policies $\pi$ results in an infinite-dimensional optimization. Moreover, the set of decision variables is non-convex for general state feedback policies \cite{goulart-06}. Instead of optimizing over all admissible policies, the suboptimal method of fixing a parameterized class such as prestablizing feedback \cite{hybrid_update_Allgower, fagiano_CDC_2012} or affine feedback policies \cite{garstka1974decision} is tractable. In prestabilizing feedback policies, the feedback gain remains constant and therefore, it is not a decision variable in the underlying optimization. Affine feedback policies are well studied along with different problem-specific options of feedback such as saturated disturbance \cite{ref:HokChaRamChaLyg-10, ref:amin-10, hokayem2009stochastic}, dropout \cite{prabhatNOLCOS2016}, innovation \cite{bosgra2003}, compensator disturbance \cite{2021_mishra_automatica}, evolving saturated disturbance \cite{2019_evolving_disturbance}, saturated innovation \cite{PDQ-LCSS, ref:Hokayem-12}, and noisy innovation \cite{2020_mishra_automatica}. Since the optimization over the general class of policies is not tractable, the optimization over feedback policies is accepted by the control community. 
	\end{enumerate}
\end{sidebar}

\section{Polynomial Chaos: An Efficient Uncertainty Propagation Framework}
Consider a model with parametric uncertainty $\theta$ with known distribution $\rho_\theta$. The objective of PCT is to describe and analyze the statistical properties of the output $y$, such as the mean, variance, skewness and kurtosis, with polynomial series, which is called a \emph{polynomial chaos expansion} (PCE). In other words, a PCE is an approximation of the implicit function between $\theta$ and $y$ with an explicit function, a linear combination of \emph{polynomial basis functions}. 

Readers are referred to \cite[\S 3]{2017_Mesbah_IJRNC} for a quick summary and to \cite{Kim_magazine_2013} for a tutorial review with historical perspectives. A survey on recent developments in PCT can be found in \cite{parametric_problems_2020}. A detailed discussion of PCT along with other methods of uncertainty propagation and quantification can be found in \cite{stochastic_finite_elements_2003, 2010_xiu, Spectral_UQ_2010, 2015_sullivan, 2018_uq}.  

\begin{figure}[b]
	\centerline{\includegraphics[width=20.0pc]{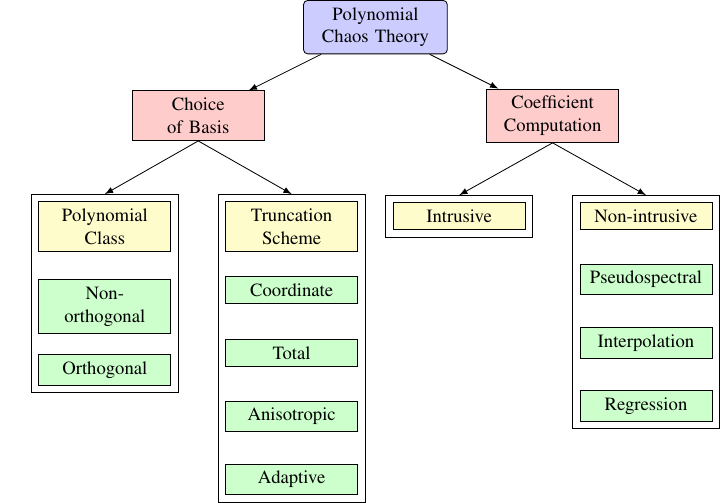}}
		\caption{Classification of PCT methods}
	\label{PCE_classification}
\end{figure}

PCT was first proposed by Norbert Wiener in 1938 \cite{wiener_1938_homogeneous}, using a Hermite polynomial basis for Gaussian random processes and its convergence was proved in 1947 \cite{cameron1947}, in what is now known as the Cameron-Martin Theorem. In the literature, the terms Wiener chaos, Wiener-Hermite chaos, and homogeneous chaos all refer to the PCE \cite{wiener_1938_homogeneous}. Any arbitrary random process with finite second-order moment can be represented by Hermite-chaos expansion \cite{wiener_1938_homogeneous} and the convergence is guaranteed by Cameron-Martin Theorem \cite{cameron1947} in the mean-square sense \cite[equation (2.38)]{Spectral_UQ_2010}. However, exponential rate of convergence with respect to the order of Hermite polynomials is not realized for processes other than Gaussian processes  \cite{2012_convergence_gPC}. Therefore, researchers have proposed other classes of polynomials. For example, Charlier polynomials are used in \cite{1972_Ogura} to represent the Poisson process. An important class of such polynomials is the Askey family \cite{1985_askey_scheme}. PCE was extended to orthogonal polynomials of the Askey scheme in \cite{Xiu_Karniadakis_PCE_tutorial} in 2002, which is also known as \emph{generalized polynomial chaos} (gPC) or \emph{Wiener-Askey polynomial chaos}.

The Wiener-Askey polynomial chaos \cite{Xiu_Karniadakis_PCE_tutorial} can be regarded as the generalization of the homogeneous polynomial chaos presented by Wiener \cite{wiener_1938_homogeneous}. In the Wiener-Askey polynomial chaos, basis functions are chosen from the Askey scheme of hypergeometric orthogonal polynomials. Each polynomial in the Askey scheme is associated with a stochastic process to which it has exponential rate of convergence in mean-square sense with respect to the order of polynomials \cite{Xiu_Karniadakis_PCE_tutorial, 2012_convergence_gPC}. Some examples of the Askey family polynomials and their corresponding probability distributions are given in Table \ref{tab:poly_distri}. For general probability distributions, the corresponding orthogonal polynomials can be obtained by the Stieltjes procedure \cite{1982_stieltjes_procedure}. 

\begin{table}[h]
	\centering
	\caption{\label{tab:poly_distri}Polynomials of the Askey scheme \cite{1985_askey_scheme} provide exponential rate of convergence for the associated distribution.}
	\begin{tblr}{
			colspec = {|c|c|},
			row{1} = {blue!15}
		}
		\hline
		 Distribution & Polynomial  \\
		\hline
	Gaussian	& Hermite \\
		\hline
	Gamma &	Laguerre  \\
		\hline
	Beta &	Jacobi \\
		\hline
	Poisson &	Charlier \\
		\hline
	Negative Binomial &	Meixner \\
		\hline
	Binomial &	Krawtchouk \\
		\hline
	Hypergeometric &	Hahn \\
		\hline
	Uniform &	Legendre \\
		\hline
	\end{tblr}  
\end{table}

Orthogonal polynomials can also be directly computed by using their properties:
\begin{enumerate}
	\item \textbf{Recurrence relation:} All orthogonal polynomials $\{\phi_n(\xi)\}$, ${n\in\Nz}$ where $n$ is the degree of the polynomial, on the real line with $\phi_{-1}(\xi) = 0$, $\phi_0(\xi) = 1$ satisfy the three-term recurrence relation 
	\begin{equation*}
		-\xi\phi_n(\xi) = b_n \phi_{n+1} (\xi) + a_n \phi_n(\xi) + c_n \phi_{n-1}(\xi), \quad n\geq 1,
	\end{equation*}
	where $b_n, c_n \neq 0$ and $\frac{c_n}{b_{n-1}} >0$. For example \cite[Appendix]{2006_sudret}, Legendre polynomials can be generated by
	\begin{equation*}
		(n+1)\phi_{n+1}(\xi) = (2n+1)\xi \phi_n(\xi) - n\phi_{n-1}(\xi)
	\end{equation*}
	and Hermite polynomials by
	\begin{equation*}
		\phi_{n+1}(\xi) = \xi \phi_n(\xi) -n\phi_{n-1}(\xi).
	\end{equation*}
	
	\item \textbf{Orthogonality relation:} The set of polynomials $\{\phi_n(\xi)\}$, belongs to the class of orthogonal polynomials if, for $n,m \in \Nz$ and support $S$ of the measure $\mu$, the relation  
	\begin{equation}\label{e:orthogonality}
		\int_S \phi_n(\xi)\phi_m(\xi) d \mu = h_n^2 \delta_{nm}
	\end{equation}
	is satisfied, where $h_n$ are scalars, and $\delta_{nm} = 1$ for $n=m$ and 0 otherwise.
	The above expression provides $\EE[\phi_n(\xi)\phi_m(\xi)]$ and is often represented as the inner product $\inprod{\phi_n}{\phi_m}$. Refer to \cite[Table B.6]{Spectral_UQ_2010} for the values of $h_n$ in some standard families of orthogonal polynomials. The set of polynomials is made orthonormal by dividing each $\phi_n(\xi)$ by $h_n$. 
	There exists a linear transformation to convert a vector of monomials to a vector with a polynomial basis and vice versa  \cite{Wan2023}.
	
	\begin{example}\label{ex:legendre}
		The uncertainty $\xi$ is uniformly distributed in the interval $[-1, 1]$, that is pdf $f_\xi(x) = 1/2$ for $x \in [-1, 1]$, otherwise zero. Table \ref{tab:poly_distri} implies that Legendre polynomials should be used. The first four normalized Legendre polynomials are given by
		\begin{align*}
			\phi_0(\xi) & = 1\\
			\phi_1(\xi) & = \sqrt{3} \xi \\
			\phi_2(\xi) & = - \frac{\sqrt{5}}{2} + \frac{3\sqrt{5}}{2} \xi^2 \\
			\phi_3(\xi) & = - \frac{3\sqrt{7}}{2}\xi +  \frac{5\sqrt{7}}{2} \xi^3. 
		\end{align*}
		It is straightforward to verify that
		\begin{align*}
			\inprod{\phi_i(\xi)}{\phi_j(\xi)} & = \EE \left[ \phi_i(\xi) \phi_j(\xi)\right] \\
			& =\int_{-1}^1 \phi_i(x) \phi_j(x) f_\xi (x) dx \\
			& = \frac{1}{2} \int_{-1}^1 \phi_i(x) \phi_j(x) dx = \begin{cases}  0 \text{ if } i \neq j \\
				1 \text{ if } i = j.
			\end{cases}
		\end{align*}
		The monomials $\xi$, $\xi^2$, $\xi^3$ can be written in terms of the orthonormal basis functions
		\begin{align*}
			\xi & = \frac{1}{\sqrt{3}} \phi_1(\xi) \\
			\xi^2 & = \frac{1}{3}\phi_0(\xi) + \frac{2}{3\sqrt{5}}\phi_2(\xi) \\
			\xi^3 & = \frac{\sqrt{3}}{5}\phi_1(\xi) + \frac{2\sqrt{5}}{7} \phi_3(\xi).
		\end{align*}
	\end{example} 
\end{enumerate}

The basic idea of polynomial chaos \cite{Xiu_Karniadakis_PCE_tutorial} is look up which of the thirteen members of the orthogonal polynomials in the Askey scheme \cite{1985_askey_scheme} to use for the probability distribution on the parameters. This approach is often referred to as generalized polynomial chaos (gPC) in the literature \cite{Xiu_Karniadakis_PCE_tutorial}. The class of Askey scheme can be further generalized by the Stieltjes procedure \cite{1982_stieltjes_procedure}, and non-orthogonal polynomials such as monic polynomials, Lagrange polynomial \cite{2005_pcm} and biorthogonal polynomial \cite{1988_biorthogonal} have been considered. 
These extensions are not discussed further here, as the probability distributions of most interest in control are included in the standard Askey scheme.

In summary, a PCE is an infinite series of orthogonal or non-orthogonal polynomials, which are called {\em basis functions}. This infinite series of polynomials is truncated for computational tractability. 
The literature on PCT can be classified (Fig. \ref{PCE_classification}) according to the choice of basis functions and the method of computation of the coefficients. The choice of basis functions can be further classified into polynomial class and truncation scheme. In literature, both non-orthogonal \cite{2005_pcm, 1988_biorthogonal} and orthogonal \cite{1985_askey_scheme} classes of polynomials have been used. Generally, the prior knowledge about the distribution of the uncertain parameters is used to decide the orthogonal polynomials as per the Table \ref{tab:poly_distri}.  Popular truncation schemes are coordinate, total \cite{Xiu_Karniadakis_PCE_tutorial}, anisotropic \cite{anisotropic_2008} and adaptive \cite{pseudo_spectral_2013}. Methods of coefficient computation are categorized into intrusive \cite{1984_galerkin} and non-intrusive methods. Non-intrusive methods include pseudospectral \cite{2007_pcm, 2010_pseudospectral}, interpolation \cite{2005_pcm} and regression \cite{2016_regression} etc. The general procedure of PCT involves three steps:
\subsection{1) Preprocessing}
In a practical problem, the uncertain parameter $\theta \in \R^{n_\theta}$ can be a correlated random variable with non-zero mean and non-identity variance. 
For the purpose of analysis, the uncertain parameter $\theta$ need to be converted into mutually independent standardized random variable $\xi$. The random variable $\xi$ is called the \emph{germ} \cite[\S 1.3]{Spectral_UQ_2010}. The continuous random variables are discretized by the Karhunen-Love expansion \cite{2003_xiu_kle}. Correlated stochastic random variables can be converted into mutually independent standardized random variables by isoprobabilistic transformations such as the Nataf or Rosenblatt transformations \cite{1996_reliability_methods,1952_rosenblatt_transformation}.

\begin{example}[\cite{Kim_magazine_2013}]
	The uncertain parameter $\theta \in \R$ is Gaussian with mean $\mu$ and variance $\sigma^2$, that is, $\theta \sim \text{Normal}(\mu, \sigma^2)$. Then $\xi = \frac{1}{\sigma} (\theta - \mu)$ is chosen to create a standardized random variable $\xi \sim \text{Normal}(0,1)$. Refer to \cite[Table 2]{Kim_magazine_2013} for transformations between the $\xi \sim \text{Normal}(0,1)$ and several common univariate distributions of $\theta$.  
\end{example}

\subsection{2) Choice of basis functions and computation of coefficients}
First consider a standardized random variable $\xi \in \R$ and a random variable $\meas \in \R$ that is an implicit function of $\xi$. A reasonable assumption is made about the germ $\xi$ and the random variable $\meas$:

\begin{assumption}[\cite{2017_Mesbah_IJRNC}]\label{as:germ_random}
	The germ $\xi$ is a random variable with finite moments of all orders (that is, $\EE[\abs{\xi}^k] < \infty $ for all $k \in \Nz$). The cumulative distribution function $F_\xi(x) = \PP (\xi \leq x)$ is continuous and uniquely defined in terms of moments of $\xi$. The random variable $\meas \in \mathcal{L}^2(\Omega, \mathcal{F}, \PP)$, where $\Omega$ is a sample space and $\PP$ is a probability measure on a sigma-algebra $\mathcal{F}$.
\end{assumption}

Since $\meas \in \mathcal{L}^2(\Omega, \mathcal{F}, \PP)$, there exists a measurable function $g: \R \rightarrow \R$ such that $\meas = g(\xi)$ due to the Doob-Dynkin Lemma \cite[page 8]{2006_probability_rao}. Assuming that $\meas$ has a finite variance, $\meas$ belongs to a Hilbert space of second-order random variables, which allows the representation
\begin{equation}
	\meas = g(\xi) = \sum_{j=0}^\infty v_j Z_j,
\end{equation}
where $\{Z_j\}_{j \in \Nz}$ forms a basis in the Hilbert space and $\{v_j\}_{j \in \Nz}$ can be interpreted as the coordinates of $\meas$. Although many choices of $\{Z_j\}$ are available, PCT usually considers an orthonormal polynomial of $\xi$ such that $Z_j = \phi_j(\xi)$. 

The canonical family of monomials $\{1, \xi, \xi^2, \ldots\}$ can be converted into orthogonal polynomials by applying the Gram-Schmidt procedure, 
\begin{equation}\label{e:gram-schmidt}
	\phi_0 = 1, \quad \phi_j = \xi^j - \sum_{k=0}^{j-1} \frac{\inprod{\xi^j}{\phi_k}}{\inprod{\phi_k}{\phi_k}} \phi_k, \quad \text{for } j = 1, 2, \ldots
\end{equation} 
Orthogonal polynomials are computed using the Gram-Schmidt procedure and the number of random variables are reduced by considering only the dominant components of the response \cite{PCE_reduced_variables_2019}.
Now $\meas  = g(\xi)$ can be represented as a weighted sum of $\phi_j(\xi)$. Let $\hat{\meas}_L$ be an approximation of $\meas$ with only $L+1$ terms in the expansion, 
\begin{equation}\label{e:truncated_pce}
	\hat{\meas}_L(\xi) \Let \sum_{i=0}^L v_i \phi_i(\xi), \quad \text{where } v_i = \frac{\inprod{g(\xi)}{\phi_i(\xi)}}{\inprod{\phi_i(\xi)}{\phi_i(\xi)}}.
\end{equation} 
The coefficient $v_i$ is called the \emph{generalized Fourier coefficient} or \emph{spectral coefficient}.
This method of coefficient computation is known as the \emph{Galerkin projection} method aka \emph{intrusive} method. The advantage of the intrusive method is that the coefficients are obtained by orthogonal projection and therefore they are optimal in the space spanned by the basis functions $\{\phi_i(\xi)\}_{i=0}^L$. Since the coefficients are computed by intruding through the model $g(\xi)$, some implicit knowledge of $g$ is required.  
The convergence and boundedness of the truncated expansion \eqref{e:truncated_pce} is assured:
\begin{theorem}[\cite{2017_Mesbah_IJRNC}]\label{th:convergence}
	Let Assumption \ref{as:germ_random} hold. Then there exists a monotonically decreasing sequence $(M_n)_{n \in \Nz}$ such that
	\begin{equation*}
		\EE [(\meas - \hat{\meas}_L)^2] \leq M_L \text{ and } \lim_{L\rightarrow \infty} M_L = 0. 
	\end{equation*}	
\end{theorem}
The first-order series expansion has been observed to be highly accurate in many control applications, but some systems have been found in which more terms are needed in the PCE. In either case,  it is recommended to confirm the accuracy of a PCE used for control design by performing \emph{a posteriori} analysis by using Monte Carlo simulation or higher order expansion \cite{Nagy_2007}.  
\begin{example}[\cite{Xiu_Karniadakis_PCE_tutorial}]
	Consider the stochastic differential equation,
	\begin{equation}\label{e:example_sde}
		\frac{dy(t)}{dt} = -\theta y, \quad y(0) = y_0.
	\end{equation}
	The decay rate $\theta$ is a continuous random variable with known distribution $\rho_\theta$. The solution $y(t) = y_0 e^{-\theta t}$ is a random variable due to the randomness in $\theta$. We first define the germ $\xi$, choose the basis function $\{ \phi_j \}$ according to the Table \ref{tab:poly_distri} and then represent $\theta$ by
	\begin{equation}\label{e:approx_theta}
		\theta \approx \sum_{j=0}^L v_j \phi_j (\xi),
	\end{equation}	
	where the coefficients $v_i$ can be computed by the inner product $\inprod{\theta}{\phi_i} = \sum_{j=0}^L v_j \inprod{\phi_j}{\phi_i}$, which gives $v_i = \inprod{\theta}{\phi_i}$. 	
	We can use orthonormal polynomial basis functions \eqref{e:truncated_pce} to represent $y$ by
	\begin{equation}\label{e:example_sde_expansion}
		y(t) \approx \sum_{i=0}^L a_i(t) \phi_i (\xi) 
	\end{equation}
	For simplicity, we will use equality sign $=$ in place of approximation sign $\approx$. Substituting the approximate expressions of $y$ \eqref{e:example_sde_expansion} and $\theta $ \eqref{e:approx_theta} into \eqref{e:example_sde} results in
	\begin{align*}
		& \frac{d}{dt} \left(  \sum_{i=0}^L a_i(t)  \phi_i \right) = -\sum_{j=0}^L v_j \phi_j \sum_{i=0}^La_i(t)\phi_i \\
		& \sum_{i=0}^L \left( \frac{d}{dt}a_i(t) \right) \phi_i = -\sum_{i=0}^L \sum_{j=0}^L v_j a_i(t)  \phi_i \phi_j.
	\end{align*}
	For each $\ell = 0, \ldots, L$, take the inner product of this equation with $\phi_\ell$ to give an ordinary differential equation for the coefficients of the PCE, 
	\begin{equation}\label{e:coefficient_evolution}
		\frac{d}{dt}a_\ell(t) = - \sum_{i=0}^L \sum_{j=0}^L v_j a_i(t)  \inprod{\phi_i \phi_j}{\phi_\ell} ; \quad  \ell = 0, \ldots, L. 
	\end{equation}
	Since the orthonormal basis functions $\{ \phi_i\}$ are chosen \emph{a priori}, the inner products $\inprod{\phi_i \phi_j}{\phi_\ell}$ can be precomputed. The expression \eqref{e:coefficient_evolution} converts a stochastic differential equation into deterministic ordinary differential equations, which can be solved to obtain an approximate solution of \eqref{e:example_sde} in a computationally efficient manner. In this example, one stochastic differential equation \eqref{e:example_sde} is converted into $L+1$ ordinary differential equations \eqref{e:coefficient_evolution}.  
\end{example}
Now consider the germ $\xi \in \R^{n_\xi}$ and assume that the components of $\xi$ are mutually independent. In this case, the polynomial series is generalized through a tensor product of univariate polynomials:
\begin{equation}\label{e:multivariate_polynomial}
	\phi_\alpha = \Pi_{i=1}^{n_\xi} \phi_{\alpha_i}(\xi_i),
\end{equation}    
where $\phi_{\alpha_i}:\R \rightarrow \R$ are univariate polynomials of order $\alpha_i$. Such polynomials can also be computed by using the Gram-Schmidt procedure by replacing $\xi$ by $\xi_i$ and $j$ by $\alpha_i$ in \eqref{e:gram-schmidt}. The multivariate polynomials $\phi_\alpha$ in \eqref{e:multivariate_polynomial} are identified by the multidimensional index $\alpha = [\alpha_i \ \cdots \ \alpha_{n_\xi}]$. The {\em order} of $\phi_\alpha$ is the $1$-norm of the vector $\alpha$, which is $\norm{\alpha}_1 = \sum_{i=1}^{n_\xi} \alpha_i$. The PCE of $\meas$ is given by
\begin{equation}\label{e:pce_all_terms}
	\meas(\xi) = \sum_{\alpha \in \N^{n_\xi}} v_\alpha \phi_\alpha(\xi).
\end{equation}
The approximation $\hat{\meas}_L$ can be found by discarding the polynomials $\phi_\alpha$ with degree larger than $d$, 
\begin{equation}\label{e:total_degree}
	\hat{\meas}_L(\xi) = \sum_{\norm{\alpha}_1 \leq d} v_\alpha \phi_\alpha(\xi) = \sum_{j=0}^L v_j \phi_j(\xi) \quad \text{where } L = \frac{(n_{\xi} + d)!}{n_{\xi} ! d!}-1.
\end{equation}  
The maximal polynomial degree $d$ has been observed to be up to $\approx 5$ for practical engineering applications \cite{power_system_2004, review_IC_PCE_2018, review_UQ_Fluid_2009, Satellite_2013, 2016_power_system, 2019_nonsmooth_PCT}. With examples of spring-mass damper system and autonomous aerial vehicles, \cite{2024_IJC} numerically demonstrates that $2-3$ degree polynomial expansion is sufficient to capture the evolution of the mean and covariance. 
The coefficient $v_j$ can be computed by Galerkin projection \eqref{e:truncated_pce} but this computation is computationally demanding for large $L$ and when governing equation $\meas = g(\xi)$ is complicated \cite{2005_pcm}. Therefore, non-intrusive methods have been proposed in literature. The non-intrusive methods include quadrature, collocation, and regression. Popular non-intrusive methods are non-intrusive spectral projection (NISP), adaptive pseudo-spectral projection \cite{winokur_2015}, hybrid least angle regression
\cite{PCE_NN_2022},
and sparse grid collocation \cite{sparse_grid_collocation_2022}. The non-intrusive methods treat the relation $\meas = g(\xi)$ as a black-box model and can be used with both orthogonal and non-orthogonal polynomials, in contrast to intrusive method. 
These methods are easy to implement but have comparatively lower accuracy and lack an optimal algorithm for choosing the collocation points on which their accuracy also depends.  

The truncation method \eqref{e:total_degree} is known as the {\em total degree method} \cite{Xiu_Karniadakis_PCE_tutorial} and is the most popular in the literature. The other popular methods are anisotropic \cite{anisotropic_2008}, adaptive \cite{pseudo_spectral_2013} and coordinate. 
Readers are referred to \cite{parametric_problems_2020} for a review of such methods. It has been observed that the coefficients of only low interaction terms are significant in applications. Therefore, low rank polynomials are proposed in \cite{2010_blatman_Sudret} in which a rank $r_{\text{max}}$ is fixed \emph{a priori} and the PCE is truncated as per the criterion $\norm{\alpha}_1 \leq d$, $\norm{\alpha}_0 \leq r_{\max}$. The truncation scheme of the total degree \cite{Xiu_Karniadakis_PCE_tutorial} is generalized by the $(q\leq 1)$-norm in \cite{adaptive_sparse_2011}, which is known as the hyperbolic truncation scheme. Another truncation scheme is based on a weighted infinity norm $\norm{D\alpha}_\infty \leq r_{\max}$, where $D$ is a diagonal matrix, and is known as the {\em coordinate degree approach}. \\

\begin{example}[\cite{2014_PCE_Sudret}]
	For $n_\xi=1, d=3$, four polynomials are needed. The Hermite polynomials computed by the corresponding differential equation and recurrence relation are $\{ 1, \xi, \frac{1}{\sqrt{2}}(\xi^2-1), \frac{1}{\sqrt{6}}(\xi^3-3\xi) \}$. For $n_\xi =2$ and $d = 3$, $10$ polynomials are needed, which are given in Table \ref{tab:hermite_poly} along with their indices $\alpha$.	
	\begin{table}[h]
	\centering
\caption{\label{tab:hermite_poly}Hermite basis for $n_\xi = 2$ and $d=3$}
	\begin{tblr}{
			colspec = {|c|c|c|},
			row{1} = {blue!15}
		}
		\hline
		$j$ & $\alpha$ & $\phi_j$ \\
		\hline
		$0$ & $(0,0)$ & $1$\\
		$1$ & $(0,1)$ & $\xi_1$ \\
		$2$ & $(1,0)$ & $\xi_2$\\
		$3$ & $(2,0)$ & $\frac{1}{\sqrt{2}}(\xi_1^2-1)$\\
		$4$ & $(1,1)$ & $\xi_1 \xi_2$\\
		$5$ & $(0,2)$ & $\frac{1}{\sqrt{2}}(\xi_2^2-1)$ \\
		$6$ & $(3,0)$ & $\frac{1}{\sqrt{6}}(\xi_1^3-3 \xi_1)$\\
		$7$ & $(2,1)$ & $\frac{1}{\sqrt{2}}(\xi_1^2-1 )\xi_2$ \\
		$8$ & $(1,2)$ & $\frac{1}{\sqrt{2}}(\xi_2^2-1 )\xi_2$ \\
		$9$ & $(0,3)$ & $\frac{1}{\sqrt{6}}(\xi_2^3-3 \xi_2)$\\
		\hline
	\end{tblr} 
\end{table}
\end{example}

Theorem \ref{th:convergence} is not applicable when the elements of $\xi \in \R^{n_\xi}$ are correlated, in which case preprocessing is required. A basis set $\{\phi_i(\xi)\}_{i \in \Nz}$ can be constructed by using the multivariate generalization of the Gram-Schmidt process \cite{2014_pce_correlated}.

We can further generalize $\meas\in \R$ to $\meas \in \R^{n_\meas}$. Consider an uncertain vector $\meas \in \R^{n_{\meas}}$ that depends on a germ $\xi \in \R^{n_\xi}$. PCT represents the $i^{\text{th}}$ element of $\meas$ in terms of the set of orthogonal basis functions $\{\phi_\alpha\}_{\alpha \in \N^{n_\xi}}$ as in \eqref{e:pce_all_terms}:
\begin{equation} \label{e:expansion}
	\meas_i(\xi) = \sum_{\alpha \in \N^{n_\xi}} v^i_{\alpha} \phi_{\alpha}(\xi) \quad \text{ for } i = 1, \ldots, n_\meas ,
\end{equation}
where each real scalar coefficient in \eqref{e:expansion} can be obtained by  $v^i_{\! \alpha} = \frac{\inprod{\meas_i}{\phi_{\! \alpha} } }{\inprod{\phi_{\! \alpha}}{\phi_{\! \alpha} } }$.
The elements of $v^i_{\! \alpha}$ can be stacked into vectors $v_{\! \alpha}$ and $\meas(\xi)$ is  approximated by
\begin{equation}\label{e:approx_PCE}
	\meas(\xi) \approx \hat{\meas}(\xi) = \sum_{\norm{\alpha}_1 \leq d} v_{\! \alpha} \phi_{\! \alpha} (\xi) = \sum_{\ell = 0}^L v_\ell \phi_\ell (\xi)
\end{equation}
by discarding the polynomials with degree larger than $d$, where $L = \frac{(n_{\xi} + d)!}{n_{\xi} ! d!}-1$, and the last equality is obtained by re-indexing the terms.

Further, $\hat{\meas}$ can be represented in a compact form
\begin{equation}\label{e:compact_PCE}
	\begin{aligned}
		\hat{\meas}(\xi) &= \bmat{v_0 & v_1 & \cdots & v_L }  \bmat{\phi_0 (\xi) \\ \phi_1(\xi) \\ \vdots \\ \phi_L(\xi) }  \teL  V^\meas \Phi(\xi) ,
	\end{aligned}
\end{equation}
where $\Phi(\xi) = [\phi_0(\xi)  \ \phi_1(\xi)  \ \cdots \ \phi_L(\xi) ]\transp \in \R^{(L+1)}$, $\phi_\ell(\xi) \in \R$, and $V^\meas \in \R^{n_\meas \times (L+1)}$. The superscript $\meas$ is used with $V^\meas$ to denote that it is a coefficient matrix in the PCE of $\meas$.

In most of the approaches, the truncated expression of the uncertain parameter $\theta$ is obtained at the beginning and problems are formulated on the basis of the expression \eqref{e:approx_PCE}. A different approach \cite{Wan2023} is to represent $\phi_j$ as a weighted sum of monomials (an expression with single non-zero term and non-negative integer exponents) with the largest degree (sum of the exponents of all variables in the expression) $d$. In particular, the $s^{\text{th}}$ monomial can be of the form 
\begin{equation} \label{e:monomial}
	\xi_s \Let \Pi_{i=1}^{n_\xi} \xi_i^{s_i} \quad 
\end{equation}
where $\sum_{i=1}^{n_\xi} s_i \leq d$.
Therefore, the problem of finding the distinct monomials is same as counting the solutions of $\sum_{i=1}^{n_\xi} s_i \leq d$, which is same as separating at most $d$ number of elements in at most $n_\xi$ number of groups.

Suppose there are a total of $L+1$  monomials present in $\theta(\xi)$ without truncation. Then $\theta$ can be represented as monomials and then each monomial can be represented by orthogonal polynomial basis functions. In particular, each $\xi_s$ for $s \in \{0, \ldots, L \}$ can be represented exactly as
\begin{equation}
	\xi_s = \sum_{k=0}^{N_s} \beta_k \phi_k(\xi),
\end{equation}
where $N_s = \sum_{i=1}^{n_\xi} s_i$ is degree of $\xi_s$. Since there is no truncation, the restriction $N_s \leq d$ is {\em not} imposed. The given random variable $\theta(\xi)$ can then be represented as a PCE by
\begin{align}
	\theta(\xi) &= \sum_{s=0}^L a_s \xi_s \notag \\
	& = \sum_{s=0}^L a_s \sum_{k=0}^{N_s} \beta_k \phi_k(\xi) = \sum_{k=0}^N c_k \phi_k(\xi), \label{e:exact_pce}
\end{align}
where $N = \max_s \{ N_s \}$.
Therefore, a random variable $\theta$ can be exactly represented with the help of orthonormal basis functions with number of terms the same as the largest degree of the monomials present in $\theta$. 

\subsection{3) Postprocessing}
Postprocessing involves moment analysis, distribution analysis, reliability analysis, and sensitivity analysis. The first two moments can be easily computed in terms of the coefficients of the expansion. The higher order moments can be obtained by combining the PCE with sampling methods such as Monte Carlo or Latin hypercube. More details about the reconstruction of complete probability distribution are provided in \cite{buehler_2016_complete_pdf}. A kernel density estimator \cite{1994_kernel_smoothing} can be used to compute nearly the exact density function. The truncated expression \eqref{e:truncated_pce} is used to compute $m^{\text{th}}$ statistical moments of $\hat{y}_L(\xi) \in \R$ by the expression
\begin{equation}\label{e:moment}
	\begin{aligned}
		\mu_m(\hat{y}) &\Let \EE [\hat{y}(\xi)^m] = \int_\Omega \hat{y}(\xi)^m \text{pdf}(\xi) \text{d}\xi \\
		& = \sum_{\ell_1 =0}^L \cdots \sum_{\ell_m = 0}^L v_{\ell_1} \cdots v_{\ell_m} \EE\left[ \phi_{\ell_1}(\xi) \cdots \phi_{\ell_m}(\xi) \right].
	\end{aligned} 
\end{equation} 
The analytical expressions of the first and second moments can be analytically computed by $\mu_1(\hat{y}) = v_0$ and $\mu_2(\hat{y}) = \sum_{\ell = 0}^L v_\ell^2$, provided $(\phi_\ell)_{\ell=0}^L$ are orthonormal. The analytical expressions of higher order moments are available when Hermite polynomials are used \cite{2014_PCE_Sudret}. Otherwise, numerical methods such as quadrature or multinomial based on Newton's multinomial theorem or recursive approaches can be applied  \cite{moment_estimation_2020}. For $y \in \R^{n_y}$, the truncated expression \eqref{e:approx_PCE} is used along with the definition of the moments of vectors instead of \eqref{e:moment}. In particular, $m^{\text{th}}$ moment of $y \in \R^{n_y}$ represents the expectation of any product of the powers of the elements of $y$ such that the sum of the powers is $m$. For example, when $n_y = 2$, the second moment of $y$ could be any one of the $\EE[y_1^2], \EE[y_1y_2]$ and $\EE[y_2^2]$. Reliability analysis is performed by defining a failure criterion known as \emph{limit state function} \cite{1996_reliability_methods} and sensitivity analysis is performed using Sobol indices \cite{2005_sensitivity_sobol}. In the context of stochastic MPC, our main focus is on moment analysis, but the other tools of PCT may be beneficial to address some associated aspects of stochastic MPC.

\begin{pullquote}
The Markovian property does not hold in the presence of time-invariant uncertainties, which makes the analysis and control design much more challenging than for control frameworks developing for parameters that are arbitrarily time varying, as discussed by \cite{2015_paulson_ti}.
\end{pullquote}

\section{Overview of Polynomial Chaos-based Stochastic MPC (PC-SMPC)} \label{s:PC-MPC}
In this section, we will provide an overview of the polynomial chaos-based stochastic MPC approaches and highlight the different techniques for linear and non-linear systems. Since the joint chance constraints (JCC) \eqref{e:jcc} play a critical role in the SOCP problem formulation, we first provide popular approaches to handle JCC  \eqref{e:jcc} in the context of SMPC.

\subsection{Chance constraints}
Chance constraints allow a certain percentage of constraints to be violated, which in turn results in a tradeoff between constraint satisfaction and performance. Since they are generally non-convex, several approximation methods \cite{2012_oldewurtel, 2009_robust_optimization, 2015_MPC_book, 2016_constraint_tightening, 2000_discretized_pdf, 2007_convex_approximation, 2009_saa, blackmore_2010_particle_chance, 2017_Mesbah_IJRNC} have been adapted by the MPC community. 
In the context of PCT, sigmoids are used in \cite{chance_constraint_sigmoid_2019} to approximate the chance constraints.

The JCC \eqref{e:jcc} $\PP[x \in \mathcal{X}] \geq 1- \beta \iff \PP[x \notin \mathcal{X}] < \beta $, which can be converted to individual chance constraints through Boole's inequality (probability of union of events is bounded by summation of the probabilities of the events) \cite{2009_convex_constraints}, $\PP[x \notin \mathcal{X}] = \PP[\cup_{\ell=1}^{n_c} \{ a_\ell^\top x > b_\ell\} ]  \leq \sum_{\ell = 1}^{n_c} \PP[a_\ell^\top x > b_\ell ]$. If we choose $\beta_\ell \geq 0$ such that $\sum_{\ell=1}^{n_c} \beta_\ell = \beta$, then 
\begin{equation}\label{e:ICC}
	\PP[a_\ell^\top x > b_\ell ] < \beta_\ell, \text{ for } \ell =1, \ldots, n_c
\end{equation}
implies satisfaction of the joint-chance constraint \eqref{e:jcc} and is called individual-chance constraint. We can use $\beta_\ell$ as decision variables in the underlying optimization problem \cite{2009_convex_constraints}. The following constraints are commonly used in place of \eqref{e:ICC}:
\begin{equation}
	\PP[a_\ell^\top x \leq b_\ell ] \geq 1-\beta_\ell, \text{ for } \ell =1, \ldots, n_c.
\end{equation}

The deterministic surrogate of the individual-chance constraints \eqref{e:ICC} is obtained by using Thm.\ \ref{th:distributionally_robust_constraints} as an application of Cantelli-Chebyshev inequality \cite{Chebyshev_inequality_1960}. This method is used in \cite{chance_constraints_2006, pseudo_spectral_collocation_22, mesbah_2014_snmpc}. We first state the Cantelli-Chebyshev inequality.
\begin{lemma}[Cantelli-Chebyshev inequality \cite{Chebyshev_inequality_1960}]\label{lem:cantelli-chebysev_inequality}
	Let $z$ be a random variable. For every $c\geq 0$, we have
	\begin{equation}\label{e:Cantelli-Chebyshev}
		\PP[z-\EE[z]\geq c] \leq \frac{\text{Var}(z)}{\text{Var}(z)+c^2}.
	\end{equation}
\end{lemma}
We can get the following result by using Lemma \ref{lem:cantelli-chebysev_inequality}.
\begin{theorem}[\S 4, \cite{2017_mesbah}]\label{th:distributionally_robust_constraints}
	Let $x$ be a random vector with mean $\EE[x]$ and co-variance $\text{var}(x)$ then
	\begin{equation}
		a_\ell^\top \EE[x] + \sqrt{\frac{1-\beta_\ell}{\beta_\ell}} \sqrt{a_\ell^\top\text{Var}(x)a_\ell} \leq b_\ell \implies \PP[a_\ell^\top x > b_\ell] < \beta_\ell.
	\end{equation}
\end{theorem}
\begin{proof}
	Let $c\geq 0$ be such that 
	\begin{equation}\label{e:constraint_with_mean}
		a_\ell^\top \EE[x] + c \leq b_\ell.
	\end{equation}
	This is immediate to notice that $\PP[a_\ell^\top x > b_\ell] \leq \PP[a_\ell^\top x \geq a_\ell^\top \EE[x] + c] \leq \frac{\text{Var}(a_\ell^\top x)}{\text{Var}(a_\ell^\top x) + c^2}$, where the last inequality is due to the Cantelli-Chebyshev inequality (Lemma \ref{lem:cantelli-chebysev_inequality}). Therefore, \eqref{e:ICC} is satisfied for
	\begin{equation*}
		\beta_\ell \geq	\frac{\text{Var}(a_\ell^\top x)}{\text{Var}(a_\ell^\top x) + c^2},
	\end{equation*}
	which gives $(1-\beta_\ell)\text{Var}(a_\ell^\top x) \leq c^2 \beta_\ell$. Therefore, we get 
	\begin{equation*}
		c \geq \sqrt{\frac{1-\beta_\ell}{\beta_\ell}} \sqrt{\text{Var}(a_\ell^\top x)}.
	\end{equation*}
	By substituting the lowest value of $c$ in \eqref{e:constraint_with_mean}, we get
	$a_\ell^\top \EE[x] + \sqrt{\frac{1-\beta_\ell}{\beta_\ell}} \sqrt{\text{Var}(a_\ell^\top x)} \leq b_\ell$.
\end{proof}
The application of Thm.  \ref{th:distributionally_robust_constraints} on individual probabilistic constraints \eqref{e:ICC} for a general class of probability distribution results into the second-order cone,   
\begin{equation}\label{e:second_order_cone}
	a_\ell^\top \EE[x] + \sqrt{\frac{1-\beta_\ell}{\beta_\ell}} \sqrt{a_\ell^\top\text{Var}(x)a_\ell} \leq b_\ell
\end{equation}
for $\ell = 1, \ldots, n_c$.
Authors in \cite{mesbah_2014_snmpc} used the truncated version of PCE of $\st$ to write the mean and variance of $\st$ in terms of the PCE coefficients by ignoring the truncation error.
The above Theorem \ref{th:distributionally_robust_constraints} is applicable for all distribution with given mean and co-variance. In some applications, the following version can be useful.
\begin{theorem}[Theorem 3.1, \cite{chance_constraints_2006}]
	Let $g(x) = a\transp x + b$ with the distribution of $[a\transp \ b]\transp$ belonging to a class $F$ with the known mean $\mu$ and known variance $\sigma$. For any $\beta \in (0,1)$, the chance constraint
	\begin{equation}
		\inf_F \  \PP [g(x) \leq 0] \geq 1-\beta, 
	\end{equation}
	is equivalent to 
	\begin{equation}
		\EE [g(x)] + \sqrt{\frac{1-\beta}{\beta}}\sqrt{\text{Var}(g(x))}  \leq 0,
	\end{equation}
	where $\text{Var}(g(x)) = \begin{bmatrix}x^\top & 1 \end{bmatrix} \sigma \begin{bmatrix}x^\top & 1 \end{bmatrix}^\top $ and $\EE[g(x)] = \mu^\top \begin{bmatrix}x^\top & 1 \end{bmatrix}^\top$.
\end{theorem}

The above approach based on Thm.\  \ref{th:distributionally_robust_constraints} can be conservative for nonlinear systems \cite{2017_mesbah}. One of the popular approaches for nonlinear systems is sample average approximation (SAA) \cite{2009_saa, blackmore_2010_particle_chance}, which considers the JCC with polytopic description of the constraint set \eqref{e:polytopic_state_constraint}. Let $c(\st_t) \Let \max_{j \in \{1, \ldots, n_c \}} a_j\transp \st_t - b_j$. SAA of JCC is provided by the empirical formula
\begin{equation*}
	\begin{aligned}
		\PP [\st_t \in \mathcal{X}] &= \PP [c(\st_t) \leq 0] = \EE [\mathbbm{1}_{\leq 0}(c(\st_t))] \\ &\approx \frac{1}{N_s}  \sum_{i=1}^{N_s} \mathbbm{1}_{\leq 0}(c(\st^{(i)}_t)), 
	\end{aligned}
\end{equation*}    
where $\mathbbm{1}$ is the indicator function and $\st_t^{(i)}$ represents $i^{\text{th}}$ sample of the state $\st_t$. The complexity of the underlying optimization significantly increases due to the non-smoothness of the indicator function and the max operator $c$, which can be addressed by employing a moment-based approximation of \eqref{e:jcc}  \cite{2017_Mesbah_IJRNC}. 

The constraint $\st_t \in \mathcal{X}$ will be satisfied with probability larger than $1-\beta$ if an ellipsoid around the mean $\st_t$ is within the set $\mathcal{X}$ with probability $1-\beta$. For an ellipsoid $\varepsilon_r \Let \{x \mid x\transp \Sigma^{-1} x \leq r^2 \}$,  it can be shown that $\PP [\st_t \in \mathcal{X}] > \PP [\st_t \in \mu_t \oplus \varepsilon_r] = \PP [(\st_t - \mu_t)\transp \Sigma_t^{-1} (\st_t - \mu_t) \leq r^2]= 1-\beta$, provided $\mu_t \oplus \varepsilon_r \subset \mathcal{X}$. Here $r$ is considered as a tuning parameter such that the condition on probability is satisfied and then the second deterministic condition can be written in terms of Mahalanobis distance by using the hyperplane representation of the set $\mathcal{X}$ \cite[Lemma 1]{2017_Mesbah_IJRNC}.   

In the remaining section, we will first present linear and nonlinear systems with full state measurements and at the end we will discuss about the filtering techniques for the incomplete and imperfect measurements.

\subsection{Linear systems with full state measurements}

Although the parameter vector $\theta$ is assumed to be time invariant, its value is uncertain and no method is available for directly measuring $\theta$. The Markovian property does not hold in the presence of time-invariant uncertainties, which makes the analysis and control design much more challenging than for control frameworks developing for parameters that are arbitrarily time varying, as discussed by \cite{2015_paulson_ti}.  The simplest form of the linear dynamics with time-invariant  $\theta$ is considered in \cite{fisher_2009_lqr, wan_CDC_2017} as a recursion:
\begin{equation}\label{e:linear_parametric}
	x_{t+1} = A(\theta)x_t + B(\theta) u_t,
\end{equation}
where the system matrices $A(\theta)$ and $B(\theta)$ depend on the uncertain parameter $\theta$ nonlinearly. The elements of $\theta$ are assumed to be mutually independent and have known probability distribution function (pdf) with finite support. The pdf quantifies the probabilistic knowledge about the value of $\theta$. 

MPC can be considered as the generalization of the Linear Quadratic Regulator (LQR) to general dynamics, cost, and constraints, whereas the LQR aims to 
design a state-feedback controller $u_t = Kx_t$ for linear dynamics to minimize the cost without constraints:
\begin{equation}\label{e:cost_infinite_horizon}
	J(x) = \EE \left [ \sum_{i=0}^{\infty} x_i \transp Q x_i + u_i \transp R u_i   \mid x_0 = x \right],
\end{equation}
for given matrices $Q \succeq 0, R \succ 0$.

The objective in \cite{fisher_2009_lqr, wan_CDC_2017} is to solve an infinite-horizon stochastic LQR problem \eqref{e:cost_infinite_horizon}.
The Foster-Lyapunov stability criterion is used to achieve mean-square stability of the system \eqref{e:linear_parametric}. The stabilizing conditions are written as a constraint in the underlying optimization. Further, \eqref{e:compact_PCE} is used to replace $x_t$, $A(\theta)$, $B(\theta)$ by polynomial basis functions and PC coefficients. By solving the conditional expectation, linear matrix inequalities (LMIs) are obtained. The truncation error is explicitly handled by a tuning parameter in \cite{wan_CDC_2017}, which provides a stability margin in the drift equation of Foster-Lyapunov stability.

A distributed MPC is proposed in \cite{pseudo_spectral_collocation_22} for the consensus control of multi-agent systems. The dynamics of the $i^{\text{th}}$ agent is given by \eqref{e:linear_parametric},
which is converted to deterministic dynamics through the gPC method and the coefficients are computed by pseudo-spectral collocation \cite{2005_pcm}. The stabilizing condition is provided in terms of a matrix inequality by assuming controllability of the matrix pair $(A(\theta), B(\theta))$ for all values of $\theta$. Multi-agent systems of the form \eqref{e:linear_parametric} with coupled chance constraints are studied in \cite{coupled_constraint_2015}. The recursive feasibility and stability both are studied in the presence of chance constraints. 

\begin{sidebar}{Example of SMPC to PC-SMPC conversion}
	\setcounter{sequation}{0}
	\renewcommand{\thesequation}{S\arabic{sequation}}
	\setcounter{stable}{0}
	\renewcommand{\thestable}{S\arabic{stable}}
	\setcounter{sfigure}{0}
	\renewcommand{\thesfigure}{S\arabic{sfigure}}
	
\sdbarinitial{L}et us consider a dynamical system \eqref{e:linear_parametric}, quadratic cost and chance constraints. The SMPC problem is given by
	\begin{equation*}
		\left .
		\begin{aligned}
			\min_{(u_{t+i\mid t})_{i=0}^{N-1}} & \quad \EE \left[ \sum_{i=0}^{N-1}  x_{t+i\mid t}^\top Q x_{t+i\mid t} + u_{t+i\mid t}^\top R u_{t+i\mid t} \right] \\
			\text{subj. to } & \quad x_{t\mid t} = x_t \\
			& \quad x_{t+i+1\mid t} = A(\theta) x_{t+i\mid t} + B(\theta) u_{t+i\mid t} \\
			& \quad \quad \quad \quad \quad \quad \text{ for } i =0, \ldots, N-1\\
			& \quad \PP \left[a_\ell^\top x_{t+i\mid t} \leq b_\ell \right] \geq 1-\beta_\ell \\
			& \quad \quad \text{ for } \ell = 1, \ldots, n_c \text{ and } i = 1, \ldots, N-1.
		\end{aligned}
		\right\}
	\end{equation*}
	Since the distribution of $\theta$ is assumed to be known, we choose the standard variable $\xi$, decide the number of terms $L+1$ in PCE and set of orthogonal polynomials $\{ \phi_0(\xi), \ldots, \phi_{L}(\xi) \}$ and represent them as $\{ \phi_0, \ldots, \phi_{L} \}$ for simplicity. The first polynomial $\phi_0(\xi) = 1$ and the remaining are according to the Table \ref{tab:poly_distri}. Readers can also refer to \cite{2025_error_bound_Automatica, Faulwasser_TAC_2022} for PCE representation by assuming the existence of the exact representation through a finite set of basis. We can pre-compute matrices $\psi_j \in \R^{(L+1)\times (L+1)}$ for $j = 0, \ldots, L$ with their $(\ell, k)^{\text{th}}$ entry 
	\begin{equation*}
		[\Psi_j]_{(\ell,k)} = \frac{\EE [\phi_j\phi_k\phi_\ell]}{\EE [\phi_\ell^2]}.
	\end{equation*}
	Let PCE representation of $A(\theta)$ and $B(\theta)$ be given by
	\begin{align*}
		A(\theta) & \approx \sum_{j=0}^L \mathbbm{A}_j \phi_j \\
		B(\theta) & \approx \sum_{j=0}^L \mathbbm{B}_j \phi_j,
	\end{align*}
	where $\mathbbm{A}_j \in \R^{n_x\times n_x}$ and $\mathbbm{B}_j  \in \R^{n_x\times n_u}$ are projections of $A$ and $B$ on $\phi_j(\xi)$, respectively. 
	We can get PCE representation of $x_{t+i\mid t}$ and $u_{t+i\mid t}$ by
	\begin{align*}
		x_{t+i\mid t} & \approx \sum_{k=0}^L \mathbbm{x}_{i,k} \phi_k \\ 
		u_{t+i\mid t} & \approx \sum_{k=0}^L \mathbbm{u}_{i,k} \phi_k.
	\end{align*}
	By substituting the PCE representations in the dynamics, we get
	\begin{equation*}
		\sum_{\ell=0}^L \mathbbm{x}_{i+1,\ell} \phi_\ell \approx \sum_{j=0}^L \mathbbm{A}_j \phi_j \sum_{k=0}^L \mathbbm{x}_{i,k} \phi_k + \sum_{j=0}^L \mathbbm{B}_j \phi_j \sum_{k=0}^L \mathbbm{u}_{i,k} \phi_k. 
	\end{equation*}
	Therefore, we can set 
	\begin{equation*}
		\mathbbm{x}_{i+1,\ell} = \sum_{j=0}^L \sum_{k=0}^L \left( \mathbbm{A}_j \mathbbm{x}_{i,k} \frac{\EE [\phi_j\phi_k \phi_\ell]}{\EE [\phi_\ell^2]} +  \mathbbm{B}_j \mathbbm{u}_{i,k} \frac{\EE [\phi_j\phi_k\phi_\ell]}{\EE [\phi_\ell^2]} \right) .
	\end{equation*}
	Let us define $\mathbbm{x}_{i} = \begin{bmatrix} \mathbbm{x}_{i, 0}^\top & \mathbbm{x}_{i,1}^\top & \ldots & \mathbbm{x}_{i,L}^\top \end{bmatrix}^\top $, $\mathbbm{u}_{i} = \begin{bmatrix} \mathbbm{u}_{i, 0}^\top & \mathbbm{u}_{i,1}^\top & \ldots & \mathbbm{u}_{i,L}^\top \end{bmatrix}^\top $. 
	Then we can get $\mathbbm{x}_{i+1} = \sum_{j=0}^L \mathbbm{A}_j\otimes [\Psi_j]_{(\ell, \cdot)} \mathbbm{x}_{i}  + \sum_{j=0}^L \mathbbm{B}_j \otimes [\Psi_j]_{(\ell, \cdot)} \mathbbm{u}_{i} $, where $[\Psi_j]_{(\ell, \cdot)}$ represents $\ell^{\text{th}}$ row of $\Psi_j$. By defining $\mathbbm{A} = \mathbbm{A}_j\otimes \Psi_j$  and $\mathbbm{B} = \mathbbm{B}_j\otimes \Psi_j$, we get
	\begin{equation*}
		\mathbbm{x}_{i+1} = \mathbbm{A}\mathbbm{x}_i + \mathbbm{B}\mathbbm{u}_i,
	\end{equation*}
	which can be used to represent evolution of the coefficients given $\mathbbm{x}_0 = \begin{bmatrix}x_{t\mid t}^\top & 0 & \ldots & 0 \end{bmatrix}^\top \in \R^{n_x (L+1)}$. 
	The chance constraint can be represented in PCE form by using \eqref{e:second_order_cone}
	\begin{equation*}
		\sqrt{ \frac{1-\beta_\ell}{\beta_\ell}}  \left( a_\ell^\top \left( \sum_{j=1}^L \mathbbm{x}_{i,j}\mathbbm{x}_{i,j}^\top \EE[\phi_j^2] \right) a_\ell \right)^{\frac{1}{2}}  + a_\ell^\top \mathbbm{x}_{i,0} \leq b_\ell,
	\end{equation*}
	for $\ell = 1, \ldots, n_c$ and $i = 1, \ldots, N-1$.
	Further we define $\varphi$ with its $(i,j)^{\text{th}}$ entry $\EE[\phi_i \phi_j]$, which is a diagonal matrix due to the orthogonality property. Now by defining $\mathbbm{Q} = Q\otimes \varphi$ and $\mathbbm{R} = R \otimes \varphi$, we can get the quadratic cost $\sum_{i=0}^{N-1} \mathbbm{x}_i^\top \mathbbm{Q} \mathbbm{x}_i + \mathbbm{u}_i^\top \mathbbm{R} \mathbbm{u}_i$. The PC-SMPC problem can be given by
	\begin{equation*}
		\left .
		\begin{aligned}
			\min & \quad \sum_{i=0}^{N-1} \mathbbm{x}_i^\top \mathbbm{Q} \mathbbm{x}_i + \mathbbm{u}_i^\top \mathbbm{R} \mathbbm{u}_i\\
			\text{subj. to } & \quad \mathbbm{x}_0 = \begin{bmatrix}x_{t\mid t}^\top & 0 & \ldots & 0 \end{bmatrix}^\top\\
			& \quad  \mathbbm{x}_{i+1} = \mathbbm{A}\mathbbm{x}_i + \mathbbm{B}\mathbbm{u}_i \text{ for } i =0, \ldots, N-1\\
			& \quad \sqrt{ \frac{1-\beta_\ell}{\beta_\ell}}  \left( a_\ell^\top \left( \sum_{j=1}^L \mathbbm{x}_{i,j}\mathbbm{x}_{i,j}^\top \EE[\phi_j^2] \right) a_\ell \right)^{\frac{1}{2}}  + \\
			& \quad a_\ell^\top \mathbbm{x}_{i,0} \leq b_\ell \text{ for }  \ell = 1, \ldots, n_c \text{ and } i = 1, \ldots, N-1.
		\end{aligned}
		\right\}
	\end{equation*}
\end{sidebar}

A linear system 
\begin{equation}\label{e:linear_parametric_additive}
	\st_{t+1} = A(\theta) \st_t + B(\theta) \control_t + E \wnoise_t
\end{equation}
with both parametric uncertainties and process noise
is considered in \cite{hybrid_update_Allgower}.
The PCE of $\st_t$, $\control_t$, $A(\theta)$, $B(\theta)$ are written and, by using the Galerkin projection, the dynamics of the PC coefficients is obtained. This control problem is converted into stochastic MPC formulation that incorporates the dynamics of the PC coefficients. Then the approach of \cite{Farina_CDC_2013} is used to approximate the chance constraints and the approach of \cite{RPI_2005} to write the terminal set. This approach relies on the pre-stabilizing feedback policies $	\control_t = \eta_t + K(\st_t - \EE[\st_t])$,
where $K$ is obtained by LQR \cite{Fisher_2008}. A simplified version of \eqref{e:linear_parametric_additive} is considered in \cite{Rolf_CDC_2015} in which the additive uncertainty $\wnoise_t$ is considered as a function of parametric uncertainty $\theta$, that is $\wnoise_t = \wnoise(\theta)$, and the existence of exact PCE is assumed. The dynamics \eqref{e:linear_parametric_additive} is considered in
\cite{paulson_stability_2015} in which predicted state-feedback policies 
\begin{equation}\label{e:policy_sf}
	\control_{t+i \mid t} = \eta_i + K_i \st_{t+i \mid t} 
\end{equation}
are considered.
In this approach, chance constraints on states are incorporated and are converted into deterministic constraints with the help of Thm.\ \ref{th:distributionally_robust_constraints}. Uncertainties are propagated through gPC and the PC coefficients are computed by Galerkin method. The Foster-Lyapunov stability is proved for the unconstrained system, which results in a bounded value function of the original constrained system. The above system \eqref{e:linear_parametric_additive} is considered in \cite{input_design_2016} for control-oriented input design. 
A continuous-time linear system with time-varying parametric uncertainties is considered in \cite{time_varying_2022}, where quadratic cost is quantified by presenting a lower bound of the cost through the zeroth-order expansion.

\subsection{Nonlinear systems with full state measurements}
The analysis of nonlinear dynamical systems is significantly different from that of linear dynamical systems \cite{2002_nonlinear_Vidyasagar}. Methods related only to PC-SMPC are discussed here, and readers are referred to \cite{2002_nonlinear_Vidyasagar} for nonlinear dynamical system analysis. 
A nonlinear discrete-time system 
\begin{equation}\label{e:nonlinear_ti}
	\st_{t+1} = f(\st_t, \control_t, \theta)
\end{equation}
is considered in \cite{fagiano_CDC_2012}. The control $\control_t$ is parameterized in terms of the policies similar to \eqref{e:policy_sf} of the form $\control_t = \eta_t + \kappa(\st_t)$,
where $\kappa$ is a precompensating controller for the system \eqref{e:nonlinear_ti}. Instead of having the constraints of the form \eqref{e:jcc}, authors have expectation-type constraints $\EE[g(\st_t)] \leq 0$. 
In order to solve the underlying constrained stochastic optimization, the state $\st_t$ and $g(\st_t)$ are represented by PCE and their coefficients are computed by weighted $\ell_2$-norm regularized regression. gPC can be applied after the uncertainty is transformed to standardized random variables, which leads to Gibbs phenomenon and affects the convergence. Therefore, the arbitrary polynomial chaos method is proposed in \cite{2018_apc} for \eqref{e:nonlinear_ti}. This method requires only statistical moments of uncertainties without the knowledge of the distribution. The arbitrary polynomial chaos method is employed in \cite{2024_arbitrary_power} to represent correlated uncertainties in power system.
A pseudo-linear description of a nonlinear system is obtained in \cite{2022_pseudo_linear} with the help of state-dependent coefficient parameterization. Both the state and control are written in terms of a gPC expansion whose coefficients are obtained by the pseudo-spectral collocation method. The controller is designed to minimize an infinite-horizon quadratic cost \eqref{e:cost_infinite_horizon} without constraints. 

A time-varying nonlinear system 
\begin{equation}\label{e:nonlinear_additive}
	\st_{t+1} = f_t(\st_t, \control_t) + \wnoise_t,
\end{equation}
without parametric uncertainties is considered in
\cite{explicit_backoff_2018},
where $f_t$ represents a nominal time-varying state transition function. This approach comprehensively considers constraint satisfaction but ignores recursive feasibility and stability. Two important aspects -- tractability with respect to the general class of policies and propagation of the process noise $\wnoise_t$ are discussed. The first aspect is addressed by fixing the class of policies and the second aspect by PCT. A PCT-based constraint tightening approach is proposed in such a way that the constraint satisfaction through nominal states implies the satisfaction of the chance constraint by the states of the system. This approach is based on the assumption that the uncertainty $\wnoise_t$ is a function of $\theta$ as in \cite{Rolf_CDC_2015}. The constraint tightening approach can be advantageous because the chance constraints are incorporated on the predicted rather than the actual states \cite{2015_paulson_ti}.

A nonlinear continuous-time system 
\begin{equation}\label{e:nonlinear_continuous}
	\dot{\st}_t = f(\st_t, \control_t, \wnoise_t, \theta)
\end{equation}
is considered in \cite{2017_Mesbah_IJRNC} in which the process noise $\wnoise_t$ is time varying. This method decouples the propagation of the probabilistic model uncertainty $\theta$ from the stochastic disturbance $\wnoise$ by using the properties of conditional probability. In particular, let $m(t) \Let \EE \left[ \st_t \mid \xi \right]$ and $V(t) \Let \text{Var} \left[ \st_t \mid \xi \right]$, where $\xi$ is a germ, which is used to represent parametric uncertainty $\theta$. The conditional mean is propagated through \eqref{e:nonlinear_continuous} and the conditional variance $V(t)$ is propagated through the linearized version of \eqref{e:nonlinear_continuous}. Since $m(t)$ and $V(t)$ are conditioned on $\xi$, they are random variables and can be represented by PCEs. Then the mean and variance of $\st_t$ are computed by using the law of total expectation and the law of total variance, respectively.

\subsection{Linear and nonlinear systems with imperfect measurements}
Any controllable LTI system in the absence of uncertainties can be exactly described with via a Hankel matrix constructed by input-output data, provided that persistency of excitation holds \cite{fundamental_lemma_2005}. This result is known as the Fundamental Lemma, which is very popular in data-driven model predictive control \cite{Koopman_behavioral_Colin, Linear_behavioral_2022, Lygeros_ECC_2019, Data_driven_stability_TAC_2020}. In \cite{Faulwasser_TAC_2022}, the Fundamental Lemma is combined with PCT for LTI systems of the form 
\begin{equation}
	\begin{aligned}
		& \st_{t+1} = A \st_t + B \control_t + \wnoise_t \\
		& \meas_t = C \st_t + D \control_t,
	\end{aligned}
\end{equation}
where $\wnoise_t \in \mathcal{L}^2$. This result is based on single-shooting algorithm, which is extended to multiple shooting in \cite{multiple_shooting_Timm_2022} to reduce the problem dimension. A tutorial review of the Fundamental Lemma and PCT can be found in \cite{2023_survey_PCE_behavoral}, which proves, under mild assumptions, that the behavior of a given $\mathcal{L}^2$ random variable is same as the coefficients of the PCE. 

To address real applications, in which the states are not perfectly measurable, PCT has been combined with state estimation techniques \cite{2009_enKF_PCT, 2010_KF_PCT, 2010_nonlinear_estimation, 2011_konda_estimation, 2013_PCE_Bayesian, bavdekar_2016_histogram}. 
The system \eqref{e:linear_parametric} is considered in \cite{output_feedback_2016} and the measurements are obtained by $\meas_t = C(\theta) \st_t + \mnoise_t$, 
where $\mnoise_t$ is measurement noise at time $t$ with known pdf. The initial state is assumed to have a known pdf but the pdf of $\theta$ is unknown. The objective is to estimate the pdf of $\theta$ and $\st_t$ given the measurement data $D_t$ up to time $t$, i.e., $D_t \Let \{ \meas_0, \ldots, \meas_t\}$. 
In this problem, the assumed pdf of the uncertain parameters is updated by using recursive parameter and state estimator based on Bayes' theorem. The basic idea of a PCE-based Bayesian filter \cite{2011_konda_estimation, 2013_PCE_Bayesian} is based on computing the PCE coefficients in such a way that the cost
\begin{equation}\label{e:moment_matching}
	\sum_{m=1}^M \abs{I_1
		- \int x_t^m\frac{\text{pdf}(\meas_t \mid \st_t) \text{pdf}(\st_t \mid D_{t-1})}{\int \text{pdf}(\meas_t \mid \st_t) \text{pdf}(\st_t \mid D_{t-1})d \st_t}dx }^2   
\end{equation}	
is minimized for $x_t \in \R$, where $I_1 = \int x_t^m\text{pdf}(x_t \mid D_t)dx$. The first integral $I_1$ in the above expression gives the $m^{\text{th}}$ moment of $x_t$ by using the posterior pdf in the left hand side of \eqref{e:Bayesian_filter} and the second integral computes the same by using the right hand side of \eqref{e:Bayesian_filter}. In order to compute PCE coefficients recursively, methods to substitute PCE of $x_t$ in \eqref{e:moment_matching} are proposed in \cite{2011_konda_estimation, 2013_PCE_Bayesian}. Different algorithms such as trust-region-reflective optimization, Levenberg-Marquardt optimization and Gauss-Newton approach can be used to solve the resulting optimization problem \cite{2013_PCE_Bayesian}. The above idea can be generalized for $x_t \in \R^{n_x}$ and for $\theta \in \R^{n_\theta}$. In case of $\theta$, we can use the expression \eqref{e:baye's_rule} to compute the first and second integrals for the moment matching as in \eqref{e:moment_matching}. In \cite{output_feedback_2016} the second integral is approximated by taking $K$ number of samples of $\hat{\theta}$ as
\begin{equation}\label{e:b_m}
	\frac{\sum_{j=1}^K (\hat{\theta}^{(j)})^m \text{pdf}(\meas_t \mid \hat{\theta}^{(j)})}{\sum_{j=1}^K \text{pdf}(\meas_t \mid \hat{\theta}^{(j)})} ,
\end{equation} 
where $\hat{\theta}$ represents PCE of $\theta\in \R$ at time $t-1$. A method to compute \eqref{e:moment_matching} with the help of histograms is proposed in \cite{bavdekar_2016_histogram}. 
The idea of the Bayesian filter is used in \cite{Imsland_Gauss_Hermite_2019, Imsland_2019, 2019_output_Imsland, Imsland_filter_2019}, with the dynamics \eqref{e:nonlinear_additive} considered with time-variant parametric uncertainty $\theta_t \in \R^{n_\theta}$. The augmented state of the system is written as $\st_t = [x^\prime_t \ \theta_t]\transp$, where $\st^\prime_t \in \R^{n_\st}$ is the state of the system. The initial state of the system is considered to be mean-zero Gaussian and the measurement equation is 
\begin{equation}
	\meas_t  = h(\st_t) + \mnoise_t,
\end{equation}
where $h:\R^{n_\st + n_\theta} \rightarrow \R^{n_\meas}$ is a measurement function and $\mnoise_t$ is zero-mean Gaussian noise. Initially, $\st_0$ is written as a PCE. At each sampling time $t+1$, the measurement $\meas_{t+1}$ is used to estimate $\st_{t+1}$ using Bayes' rule and the PCE representation of $\st_{t+1}$ is updated accordingly. 

A continuous-time stable linear differential algebraic system 
\begin{equation}\label{e:descriptor}
	M(\theta) \dot{\st} = A(\theta)\st + B(\theta)\control + \wnoise(\theta)
\end{equation}  
is considered in \cite{2014_fast_braatz} in which $\st$ consists of both differential (states whose derivatives appear in the vector $M(\theta)\dot{\st}$) and algebraic (remaining states and measurement $\meas$) states. Each element of $M(\theta)$, $A(\theta)$, $B(\theta)$, $\wnoise(\theta)$,  and $\st$ is represented by \eqref{e:approx_PCE}. The explicit dependence of $M$, $A$, $B$, and $\wnoise$ on $\theta$ is assumed to be known \emph{a priori} and so their corresponding PCE coefficients can be precomputed by Galerkin projection. Further, the substitution of PCE of each element in \eqref{e:descriptor} and application of the Galerkin projection one more time provides another set of discrete algebraic equations without parametric uncertainty $\theta$. In many engineering applications, the number of states $n_x$ is large and PCE system has $(L+1)n_\st$ states. For example, an end-to-end continuous pharmaceutical manufacturing process has approximately $8000$ states \cite{2014_fast_braatz}. Such systems can be handled by an input-output framework based on a finite step response model \cite{2013_Ikonen,2014_fast_braatz}. This formulation results in offset which can be removed by modifying the penalty of the variance of the output in the cost function \cite{2019_von_andrien_offset_free}. The approach \cite{2019_von_andrien_offset_free} is extended in \cite{2020_von_unstable} for unstable systems by using a prestabilizing output feedback.   

\section{Engineering Applications of PC-SMPC}
This section describes some engineering applications of PC-SMPC and highlights which technical approach 
was used. In addition, implementation issues and the software packages are discussed. A non-exhaustive list of some engineering applications is provided in Table \ref{tab:eng_applications} and some popular toolboxes are enlisted in Table \ref{tab:toolbox}. 

\begin{table}[h]
  \centering
\caption{\label{tab:eng_applications}Engineering applications of PC-SMPC}  
  \begin{tblr}{
    colspec = {|m{2cm}|m{0.8cm}|m{1.8cm}|m{1.8cm}|},
    row{1} = {blue!15},
  }
    \hline
      Application & State dim. & Uncertainty distribution & Implementation  \\
    \hline
         Autonomous driving  \cite{Autonomous_navigation_Jones_2021} & 6 & uniform & numerical \\
    \hline   
    Motion control \cite{2023_av} & 7 & Gaussian & numerical\\
    \hline 
   Trajectory optimization \cite{motion_planning_2023} & 13 & Gaussian & spacecraft simulator \\
    \hline 
    Haptic shared control \cite{haptic_20} & 8 & Gaussian & numerical \\
    \hline 
    optimal charging of battery pack \cite{EV_battery_22} & 12 & Gaussian & numerical\\
    \hline
    Propulsion system \cite{propulsion_system_22} & 4 & Gaussian & numerical \\
    \hline
    Reservoir management \cite{Reservoir_22} & 2& Gaussian \& uniform & numerical\\
    \hline  
    Fault diagnosis \cite{mesbah_2014_diagnosis} & 3 & Gaussian & numerical \\
    \hline
    Pharmaceutical manufacturing \cite{2018_pharmaceutical} & 8000 & Gaussian & numerical\\
    \hline    
  \end{tblr}
\end{table}

Autonomous driving applications require road safety through trajectory planning even if the parameters of driving vehicle are not precisely known. A Dubin's car model is considered in \cite{Autonomous_navigation_Jones_2021}, which has the dynamics of the form \eqref{e:nonlinear_continuous}. This continuous-time nonlinear system is represented by gPC and its coefficients are computed by pseudospectral collocation method. The chance constraints are converted to deterministic constraints via Thm.\  \ref{th:distributionally_robust_constraints} and the control has a pre-stabilizing feedback component. The authors used the \texttt{C++}-based toolbox PolyMPC \cite{2020_polympc} to simulate the system. In \cite{2023_av}, the trajectory tracking problem for autonomous vehicles is studied and ACADOS library \cite{2022_acados} is used to generate the \texttt{C} code. Motion planning for robotic systems under uncertainties is considered in \cite{motion_planning_2023} by converting the continuous-time chance-constrained stochastic optimal control to its deterministic surrogate through gPC \cite{stochastic_finite_elements_2003}. The motivation of using gPC is that the chance-constraints are non-convex in moment space but become convex in gPC space, which allows the use of \cite{swarms_2014} to compute the solution.  
A haptic-shared control framework is presented in \cite{haptic_20} in which a human driver and automation system share a steering wheel. A cost function is constructed to reduce the disagreement between human and automation system. PCE is employed to propagate stiffness/damping uncertainties in human's biomechanics. A numerical case-study involving F-16 aircraft is presented in \cite{2024_F16}.

A stochastic MPC formulation for the optimal charging of a Li-ion battery pack has been proposed \cite{EV_battery_22}. The system dynamics is governed by a control-oriented electrochemical model given by a DAE system,  
uncertainties are propagated through gPC, and constraints are softened by Thm.\  \ref{th:distributionally_robust_constraints}. L$_2$-norm regularized optimization  \cite{Stochastic_simulation_Khammas_2012} and regularized least-squares regression \cite{fagiano_CDC_2012} were applied.  
Polynomials up to degree two were considered which, in the presence of $44$ parameters, resulted in $1035$ terms in the PCE.

A kinetic model to describe the dynamics of the bioproduction of succinic acid from glycerol is developed in \cite{bioreactor_22} in which PCT is used for uncertainty propagation through dynamic black-box simulator and to determine the optimal control policy to maximize the yield of succinic acid while reducing the concentration of the side products.
In a biochemical setting, performing an experiment is costly, and control and sampling schedules are designed to maximize the information content by optimal model-based experiment design. The sensitivity-based approach \cite{2004_oed} and PCT-based approach for the optimal experiment design are compared in \cite{comparison_PCE_sensitivity}. Two case studies, namely the Lotka--Volterra predator prey model and a jacketed tubular reactor, were implemented using \texttt{CasADi} in \cite{casadi}. PCT was observed to perform well in terms of the information content but violated more constraints with respect to \cite{2004_oed}.

A multi-fidelity approach for efficient uncertainty quantification that combines gPC with Monte Carlo has been proposed \cite{propulsion_system_22}. Mean, variance, and skewness of axle shaft torque are computed for an automotive propulsion system during the vehicle launch. PC-SMPC has been used in petroleum reservoir simulation to balance economics and safety  \cite{Reservoir_22}, with an objective to maximize the oil production rate while tracking the factor associated with safety.  
A fault detection and isolation problem is considered in \cite{mesbah_2014_diagnosis} in which probability distributions of the output signal with respect to all fault scenarios are required to compare the respective Hellinger distance. Nominal and all faulty systems with the parametric uncertainties are represented in terms of PCEs, with their coefficients computed by the collocation method and Gaussian quadrature rule.

Continuous manufacturing processes, in which the input materials are continuously fed into the system and processed materials are continuously removed, are more economical with respect to batch processes. MPC approaches for continuous manufacturing have been reviewed elsewhere \cite{mpc_manufacturing_2021}. PC-SMPC has been demonstrated in a commercial simulator for the end-to-end integrated continuous manufacturing of a pharmaceutical  \cite{2018_pharmaceutical}. To obtain fast computation, an input-output PC-SMPC framework was formulated that integrated PCT and Quadratic Dynamic Matrix Control. The  process model had nearly $8000$ states but the online computational cost was very low ($<1$ s) by formulating the on-line calculations to have cost that is independent of the number of states. Off-line calculations were used to construct the input-output operators needed for online calculation, so the online computation cost was only a function of the number of manipulated variables and the control and prediction horizons.

Explicit MPC is an MPC formulation designed with the objective of reducing online computational cost by partitioning the state-space and providing explicit control laws for each partition that are selected online using a lookup table. The onboard memory depends on the number of partitions which grows exponentially with the number of constraints and the control and prediction horizons. To reduce the memory requirements, the explicit control laws have been approximated by a Legendre polynomial series expansion constructed by applying Galerkin projection to the KKT conditions \cite{control_law_approximation_2019}. Incorporating PCT within such as a formulation, to explicitly address time-invariant parametric uncertainties, would be expected to be straightforward, given the common basis functions.

A model order reduction framework has been proposed to integrate PCT and NN for hybrid simulation \cite{PCE_NN_2022}. Hybrid simulation studies the response of a system while including hardware in the loop. In order to get high-fidelity results, real-time implementation is required and  dimensionality reduction can be beneficial to reducing online computational cost. The \texttt{matlab}-based software package \cite{2014_uqlab} was used to simulate the numerical sub-system by PCT.

\begin{table}[h]
  \centering
  \caption{\label{tab:toolbox}Popular toolboxes for PCT and PC-MPC}
  \begin{tblr}{
    colspec = {|c|c|c|},
    row{1} = {blue!15},
    % column{3} = {teal7},
    % cell{2}{3} = {yellow7},
  }
    \hline
      Toolbox &  Language & Reference \\
    \hline
    UQLab & \texttt{MATLAB} & \cite{2014_uqlab}\\
    \hline 
        PoCET & \texttt{MATLAB} & \cite{pocet_2020}\\
        \hline
        uncertainpy & \texttt{python} & \cite{uncertainpy_2018}\\
  \hline  
  pygpc & \texttt{python} & \cite{pygpc_2020} \\
  \hline
  PolyChaos & \texttt{Julia} & \cite{PolyChaos_2020}\\
  \hline 
  Chaospy & \texttt{python} & \cite{2015_chaospy}\\
  \hline
  PolyMPC & \texttt{C++} & \cite{2020_polympc} \\
  \hline
  \end{tblr}
\end{table}

\section{Conclusion and future outlook}
The objective of this article is to provide a tutorial introduction of polynomial chaos theory-based model predictive control that describes all key considerations of such formulations, while reviewing the literature which is summarized in tabular format in Table \ref{tab:lit_summary}. 
Although PCT facilitates the propagation of probabilistic uncertainties in time-invariant parameters to the states, outputs, and constraints, the truncation error affects the existing methods of stochastic MPC designed to ensure stability, recursive feasibility, and convex approximation of chance constraints. Readers interested more generally in open problems in control theory are referred to \cite{2012_open_problems}.
\begin{table*}[h]
\centering
\caption{\label{tab:lit_summary}Literature Summary}
\begin{tabular}{|m{1.5cm}|m{2.5cm}|m{5cm}|}    
    \hline
    \rowcolor{blue!15}
    & Comment & Literature  \\
    \hline
    
    \multirow{3}{*}{Basics} & MPC & \cite{2017_Rawlings_book}\\
    \cline{2-3} 
    & PCT & \cite{stochastic_finite_elements_2003, 2010_xiu, Spectral_UQ_2010, 2015_sullivan, 2018_uq} \\
    \cline{2-3}
    & Nonlinear system analysis & \cite{2002_nonlinear_Vidyasagar} \\
    \cline{2-3}        
    \hline
        
    \multirow{2}{*}{Review} & MPC & \cite{2016_review_SMPC, mpc_manufacturing_2021} \\
    \cline{2-3}
    & PCT & \cite{Kim_magazine_2013, parametric_problems_2020} \\ 
    \cline{2-3}
    & Behavioral approach & \cite{2023_survey_PCE_behavoral} \\                     
    \hline 
               
    \multirow{4}{*}{PC-SMPC} 
    & Linear systems  & \cite{fisher_2009_lqr, wan_CDC_2017, pseudo_spectral_collocation_22, coupled_constraint_2015, hybrid_update_Allgower, paulson_stability_2015, input_design_2016, time_varying_2022} \\
    \cline{2-3} 
              
    & Nonlinear systems & \cite{fagiano_CDC_2012, 2018_apc, 2022_pseudo_linear, explicit_backoff_2018, 2017_Mesbah_IJRNC} \\
    \cline{2-3}
              
    & Imperfect measurements & \cite{Faulwasser_TAC_2022, multiple_shooting_Timm_2022, 2009_enKF_PCT, 2010_KF_PCT, 2010_nonlinear_estimation, 2011_konda_estimation, 2013_PCE_Bayesian, bavdekar_2016_histogram, output_feedback_2016, Imsland_Gauss_Hermite_2019, Imsland_2019, 2019_output_Imsland, Imsland_filter_2019, 2014_fast_braatz, 2019_von_andrien_offset_free, 2020_von_unstable} \\
    \cline{2-3}
              
    & Chance constraints & \cite{chance_constraint_sigmoid_2019, chance_constraints_2006, pseudo_spectral_collocation_22, mesbah_2014_snmpc} \\
    \hline
              
    \multirow{2}{*}{Applications} 
    & PCT & \cite{power_system_2004, review_IC_PCE_2018, review_UQ_Fluid_2009, Satellite_2013, 2016_power_system, 2019_nonsmooth_PCT} \\
    \cline{2-3}
    & PC-SMPC & \cite{Autonomous_navigation_Jones_2021, motion_planning_2023, haptic_20, EV_battery_22, bioreactor_22, propulsion_system_22, Reservoir_22, mesbah_2014_diagnosis, nonlinear_surrogate_PCE, 2018_pharmaceutical}\\
    \hline   
\end{tabular}
\end{table*}

This section highlights some interesting directions for research to further progress of stochastic MPC in the framework of PCT. First, it is useful to make some more general comments on stochastic MPC. On one hand, stochastic MPC algorithms have been developed that are easy to implement, have low online computational cost, and have demonstrated high closed-loop performance in nontrivial dynamical systems with large number of uncertain parameters and states but lack  theoretical guarantees. On the other hand, some stochastic MPC algorithms offer theoretical guarantees of recursive feasibility and stability, but their suitability for real-world applications is limited due to only minor improvements over simpler methods at the cost of significantly higher computational complexity \cite{2015_mayne_robust_stochastic}. As for stochastic MPC, some promising directions in control would be to develop PC-SMPC algorithms that have all of the capabilities of the most practical algorithms while providing theoretical guarantees currently available only for the algorithms that have limited practical utility \cite{2015_paulson_ti}. 

Below are more detailed discussion of the promising future directions in PC-SMPC, with some related comments with regard to stochastic MPC more generally.

\subsection{Sparsity-promoting methods for PC-SMPC}

A high-order PCT has a large number of terms for systems with a large number of parameters. It is observed that the non-zero coefficients of PCE often form a sparse subset. As such, the cost can be reduced, typically by many orders of magnitude, by choosing a sparse basis instead of computing terms with negligible effect. Adaptive algorithms have been proposed and demonstrated in \cite{2010_blatman_Sudret, adaptive_sparse_2011} using least-angle regression techniques \cite{2004_lar}. Readers are referred to \cite{2011_sudret_pca, 2013_sudret_sparse} for vector-valued models. A sparse Gauss-Hermite sampling rule for efficient scenario-based NMPC has been proposed  \cite{Imsland_Gauss_Hermite_2019}. Sparse PCT and Krigging methods are used to construct a surrogate of a population balance model in \cite{2020_surrogate_modeling}, where sparsity is induced by the least-angle regression method. This method has been  used for uncertainty quantification and sensitivity analysis of a simulated mars life support system \cite{2020_mars_life_support}. The idea of basis-adaptive sparse regression is used to develop a non-smooth PCT in \cite{2019_nonsmooth_PCT} for dynamic flux balance analysis. An extensive review on sparse PCT is available \cite{Sparse_PCE_2022} that includes benchmarks of nonintrusive regression-based algorithms. A promising direction is to explore the use of these sparsity-promoting methods in the context of PC-SMPC.

\subsection{Input-output model for PC-SMPC}

The implementation of MPC for a large number of states is difficult for most MPC formulations and that of state space-based PC-SMPC is even more difficult. The implementation for high state dimension is easy for input-output formulations such as using a finite step response for the process and the PC coefficients \cite{2014_fast_braatz, 2019_von_andrien_offset_free, 2020_von_unstable}. PC-SMPC for autoregressive models along with sparsification algorithms such as elastic net can be an interesting direction to handle parametric uncertainties in high-dimensional systems. This approach would integrate PCT into existing nonlinear input-output MPC formulations   \cite{2023_NARX_io_model} to enable them to explicitly account for time-invariant probabilistic parametric uncertainties. CLOT (Combined L-One and Two) norm minimization \cite{2020_clot} is a related alternative method to elastic net that had better performance in some control case studies.

\subsection{Time-varying basis for PC-SMPC}

In MPC with a long time horizon, the PCE of the random variable with fixed basis may lose accuracy for systems that are oscillatory or unstable \cite{Bonnaire}. In those situations, time-varying basis functions may be helpful \cite{2010_time_dependent_pc}. 

\subsection{Analysis of truncation error in  PC-SMPC design}

Most of the literature in stochastic MPC ignores the truncation error in the analysis (recursive feasibility and stability) or assumes the existence of an exact finite expansion \cite{Faulwasser_TAC_2022, hybrid_update_Allgower}. 
While some PCT-based control algorithms have been developed that provide stability guarantees that explicitly account for truncation error (the first being \cite{Lucia_ACC_2017}), those developments were for the design of time-invariant controllers without consideration of constraints. The effect of truncation error and its bounds have been discussed in some detail in the literature \cite{error_bound_2017, truncation_error_2018}. Even if the PCE-transformed system is stable, the actual closed-loop system can be unstable due to truncation error. 

Several research questions are explored in \cite{Wan2023, Wan2021}, including strategies for transformation to minimize the truncation error in PCE transformations, incorporating truncation error considerations into controller design, and ensuring stability of the closed-loop system by demonstrating stability of the PCE transformed system. Various approaches have been attempted to address these questions. For instance, \cite{wan_CDC_2017} focuses on transforming the cost function of the optimization problem rather than the system dynamics themselves. In \cite{roa_2020}, the relationship between the actual and PCE-transformed systems is examined in terms of their respective regions of attraction (ROAs), along with a numerical method for estimating an inner ROA. Additionally, \cite{wan_CDC_2017} explicitly deals with truncation error by introducing a tuning parameter that provides a stability margin in the drift equation of Foster-Lyapunov stability \cite{1993_stability_Meyn_Tweedie}. Moreover, \cite{Wan2023} explores the transformation of closed-loop states, resulting in reduced truncation error.

A detailed discussion of these problems in the context of PC-SMPC is essential for the further advancements of theory. The PCE of the state $\st_t$ is valid as per the Assumption \ref{as:germ_random} when $\st_t \in \mathcal{L}^2$. This assumption is often overlooked in the control literature. 

\subsection{Recursive feasibility of PC-SMPC}

Recursive feasibility in the context of stochastic MPC has received attention in several works  \cite{2009_primbs_recusrsive_feasibility, policy_2008, 2006_chance_constraint_approximation, 2017_mesbah, 2021_constraint_violation, 2021_recursive_feasibility, 2014_korda_average_constraint_violation, 2021_discounted_constraint}.  
Most of these methods are not considered in the context of PC-SMPC. In any MPC-based controller design, the guarantee of recursive feasibility is necessary; otherwise the constraints should be reformulated to be soft (if the physical system allows such softening) or some backup controller or recovery strategy should be provided.

\subsection{Symbolic toolboxes for PC-SMPC}

A symbolic method to automatically create PCE representations of dynamical systems is presented in \cite{von_ACC_2020}, which relies on the symbolic toolbox of \texttt{MATLAB}. A nonlinear continuous-time state-space system is first represented symbolically, and the desired PCE order and the distribution of the parametric uncertainty $\theta$ are specified. Then two paths are taken to obtain a linear PCE system. In the first path, the state-space system is transformed into the PCE system and then linearized. In the second path, the nonlinear state-space system is first linearized and then converted into the linear PCE system. The symbolic representation allows to use the Galerkin method easily to compute the coefficient which in turn results in zero aliasing error. This approach demonstrates a promising way for PC-SMPC design but only implementable for moderate state dimensions $n_\st$ with a low number of uncertain parameters $n_\theta$. The symbolic arithmetic calculations in \cite{von_ACC_2020} are exact for polynomial systems. Further development of this method with sparse PCT and generalization for non-polynomial systems with alternative approaches or approximations \cite{2019_Harinath_polynomial} may speed up the progress in designing and testing of PCT-based products. The Weirstrass approximation theorem can be applied to non-polynomial continuous functions on compact sets, then their PCE is obtained. It is well known that a global approximation can be oscillatory even with high-degree polynomials. Therefore, it may be useful for some problems to first divide the whole interval in subintervals, and then approximate the nonlinear function with the help of low-degree polynomials for each sub-interval \cite[Chapter 11]{numerical_analysis_2003}.

\begin{sidebar}{Learning-based MPC}
	\setcounter{sequation}{0}
	\renewcommand{\thesequation}{S\arabic{sequation}}
	\setcounter{stable}{0}
	\renewcommand{\thestable}{S\arabic{stable}}
	\setcounter{sfigure}{0}
	\renewcommand{\thesfigure}{S\arabic{sfigure}}
	% \subsection{Learning-based MPC}
	\sdbarinitial{L}earning-based MPC has attracted the attention of researchers due to the constraint-handling capability of MPC to ensure safety in learning-based techniques for safety critical applications \cite{lbmpc_linear, PG20, LMPC_Borrelli, 2023_unmatched, learning_MPC_Mesbah}. The readers are referred to \cite{survey_LMPC, safe_learning_survey_2022} for some recent surveys on safe learning. The capabilities of PCT in the context of learning-based MPC are not explored yet. Only a handful literature is available where PCT is employed along with some learning-based techniques.

	Application of PCT for regression problems is investigated in \cite{2019_PC_ML} and it is demonstrated that the accuracy of PCT is comparable to other machine learning models such as neural networks and support vector machines. Advantages of PCT over popular machine learning methods include less parameter tuning, retaining the statistical properties of the output, interpretability of the model, portability to embedded devices, and its sample efficiency.  
	Most of the methods of Physics-Informed Neural Network (PINN) are available for solving only PDEs. In order to solve the stochastic PDE, a quick quantification of the impact of the parametric uncertainty is needed. In such situations, PCT can be employed with some modifications \cite{Newman_Neurips_2021}.

	A tutorial on PCT from a statistician's perspective is presented in \cite{OHagan_SIAM_2013} which includes a comparison of PCT with respect to Gaussian Process (GP). GPs are well known for approximating the function locally while PCT is known for global approximation.
	GP Krigging and PCT are presented for meta-modeling in \cite{2023_PCE_GP} and are combined by representing the mean of GP by PCT. A similar idea also appears in \cite{Imsland_2021}. A robust nonlinear model predictive control scheme with PCT and recurrent neural network is developed in \cite{2024_PC_RNN} by using PCT for predicting the statistical moments, which are projected to a low dimensional space and then recurrent neural networks are used to learn the reduced dynamics of moments. An idea of deep PCT has been introduced in \cite{2025_deep_PCE} with an intention to support analytical uncertainty quantification by simple forward passes and to overcome the curse of dimensionality of PCT. A semi-supervised learning approach is proposed \cite{2023_deep_aPC} in which properties of adaptive PCT are used to assist training of deep neural network in the presence of small amount of labeled data. Further, a mathematical formulation of kernel in deep artificial neural network with PCT is introduced in \cite{2023_oladyshkin} along with a MATLAB toolbox \cite{2025_apc_matlab_toolbox}.
\end{sidebar}

 \section{acknowledgment}
This research was supported by the U.S. Food and Drug Administration under the FDA BAA-22-00123 program, Award Number 75F40122C00200.

\section{Author Information}

\begin{IEEEbiography}{{P}rabhat K. Mishra}{\,}(pkmishra@ai.iitkgp.ac.in)  is an Assistant Professor at the Indian Institute of Technology Kharagpur (IITKgp) where he does research at the intersection of control theory and artificial intelligence for safety-critical applications and cyber-physical systems. 
	He received a Ph.D. from the Indian Institute of Technology Bombay (IITB) and a Swiss Government Excellence Scholarship to study at École polytechnique fédérale de Lausanne (EPFL) for a year. He was a Postdoctoral Associate at the Massachusetts Institute of Technology (MIT) and a Postdoctoral Research Associate at the University of Illinois at Urbana-Champaign (UIUC) before moving to IITKgp.  
\end{IEEEbiography}

\begin{IEEEbiography}{{J}oel A. Paulson}{\,} received the B.S. degree with highest honors in chemical engineering from the University of Texas at Austin (UT Austin), Austin, TX, USA in 2011 and the M.S.CEP. and Ph.D. degrees in chemical engineering from the Massachusetts Institute of Technology (MIT), Cambridge, MA, USA in 2013 and 2016, respectively. From 2016 to 2019, he was a postdoctoral scholar in systems and control theory at the University of California, Berkeley (UC Berkeley), Berkeley, CA, USA. In 2019, he began his faculty career in the Department of Chemical and Biomolecular Engineering at The Ohio State University (OSU), Columbus, OH, USA where he was the H.C. ``Slip'' Slider Assistant Professor. In 2025, he started as the Battist Associate Professor of Chemical and Biological Engineering at the University of Wisconsin-Madison (UW-Madison), Madison, WI, USA. His research interests include data-driven optimization and control of complex systems under uncertainty, with applications in chemistry, biology, energy, and next-generation manufacturing, as well as the mathematical foundations of learning. He was the recipient of the Best Application Paper Prize from the 2020 International Federation of Automatic Control (IFAC) World Congress, the 2022 Lumley Research Award from OSU, and the 2025 David C. McCarthy Engineering Teaching Award from OSU. He received the National Science Foundation CAREER award in 2023 and the 35 under 35 award from the American Institute of Chemical Engineers (AIChE) in 2023. He was also selected as a member of the 2025 Class of Influential Researchers by the American Chemical Society (ACS) Industrial \& Engineering Chemistry Research journal in 2025. He is a Member of IEEE.
\end{IEEEbiography}

\begin{IEEEbiography}{{R}ichard D. Braatz}{\,} is the Edwin R. Gilliland Professor at the Massachusetts Institute of Technology (MIT) where he does research in control theory and its applications to advanced manufacturing. He received an M.S. and Ph.D. from the California Institute of Technology and was a Professor at the University of Illinois at Urbana-Champaign and a Visiting Scholar at Harvard University before moving to MIT. Recognitions include the AACC Donald P. Eckman Award and the IEEE CSS Antonio Ruberti Young Researcher Prize. He is a Fellow of IEEE 	and IFAC and a member of the U.S. National Academy of Engineering.
\end{IEEEbiography}

\bibliographystyle{IEEEtran}       
\bibliography{smpc_poly_chaos.bib} 

@article{2011_TAC_Lygeros,
  title={Stochastic receding horizon control with bounded control inputs: A vector space approach},
  author={Chatterjee, D. and Hokayem, P. and Lygeros, J.},
  journal={IEEE Trans. on Autom. Control},
  volume={56},
  number={11},
  pages={2704--2710},
  year={2011}
}

@incollection{2018_Constant_uncertainty,
  title={Uncertain systems: Time-varying versus time-invariant uncertainties},
  author={Blanchini, F. and Colaneri, P.},
  booktitle={Uncertainty in Complex Networked Systems: In Honor of {Roberto Tempo}},
  pages={3--91},
  year={2018},
  publisher={Springer}
}

@article{2005_linear_tube,
  title={Robust model predictive control of constrained linear systems with bounded disturbances},
  author={Mayne, D. Q. and Seron, M. M. and Rakovi{\'c}, S. V.},
  journal={Automatica},
  volume={41},
  number={2},
  pages={219--224},
  year={2005},
  publisher={Elsevier}
}

@article{2023_oladyshkin,
  title={The deep arbitrary polynomial chaos neural network or how Deep Artificial Neural Networks could benefit from data-driven homogeneous chaos theory},
  author={Oladyshkin, S. and Praditia, T. and Kroeker, I. and Mohammadi, F. and Nowak, W. and Otte, S.},
  journal={Neural Networks},
  volume={166},
  pages={85--104},
  year={2023},
  publisher={Elsevier}
}

@misc{2025_apc_matlab_toolbox,
  title={{aPC Matlab} toolbox: data-driven arbitrary polynomial chaos},
  author={Oladyshkin, S.},
  year={2025},
 url = {https://www.mathworks.com/matlabcentral/fileexchange/72014-apc-matlab-toolbox-data-driven-arbitrary-polynomial-chaos},
 publisher = {MATLAB Central File Exchange}
}

@article{2023_deep_aPC,
  title={Deep adaptive arbitrary polynomial chaos expansion: A mini-data-driven semi-supervised method for uncertainty quantification},
  author={Yao, W. and Zheng, X. and Zhang, J. and Wang, N. and Tang, G.},
  journal={Reliability Engineering \& System Safety},
  volume={229},
  pages={108813},
  year={2023},
  publisher={Elsevier}
}

@inproceedings{2025_deep_PCE,
  title={Deep Polynomial Chaos Expansion},
  author={Exenberger, J. and Ranftl, S. and Peharz, R.},
  booktitle = {Eighth Workshop on Tractable Probabilistic Modeling},
  year={2025}
}

@article{2024_PC_RNN,
  title={Robust model predictive control for large-scale distributed parameter systems under uncertainty},
  author={Tao, M. and Zacharopoulos, I. and Theodoropoulos, C.},
  journal={arXiv preprint arXiv:2410.12398},
  year={2024}
}

@article{2019_PC_ML,
  title={Data-driven polynomial chaos expansion for machine learning regression},
  author={Torre, E. and Marelli, S. and Embrechts, P. and Sudret, B.},
  journal={Journal of Computational Physics},
  volume={388},
  pages={601--623},
  year={2019},
  publisher={Elsevier}
}

@article{2018_Automatica_Bitmead,
  title={Stochastic output-feedback model predictive control},
  author={Sehr, M. A. and Bitmead, R. R.},
  journal={Automatica},
  volume={94},
  pages={315--323},
  year={2018},
  publisher={Elsevier}
}

@article{2025_error_bound_Automatica,
  title={A polynomial chaos approach to stochastic {LQ} optimal control: Error bounds and infinite-horizon results},
  author={Ou, R. and Schie{\ss}l, J. and Baumann, M. H. and Gr{\"u}ne, L. and Faulwasser, T.},
  journal={Automatica},
  volume={174},
  pages={112117},
  year={2025},
  publisher={Elsevier}
}

@inproceedings{2009_scenario_Bemporad,
  title={Scenario-based model predictive control of stochastic constrained linear systems},
  author={Bernardini, D. and Bemporad, A.},
  booktitle={48h IEEE Conf. on Decision and Control held jointly with 28th Chinese Control Conf.},
  pages={6333--6338},
  year={2009},
  organization={IEEE}
}

@article{2024_IJC,
  title={Differential flatness based design of robust controllers using polynomial chaos for linear systems},
  author={Ogunbodede, O. and Singh, T.},
  journal={International Journal of Control},
  volume={97},
  number={8},
  pages={1687--1703},
  year={2024},
  publisher={Taylor \& Francis}
}

@article{2023_ti,
  title={Stochastic optimal linear control for generalized cost functions with time-invariant stochastic parameters},
  author={Ito, Y. and Fujimoto, K. and Tadokoro, Y.},
  journal={IEEE Trans. on Cybernetics},
  volume={54},
  number={6},
  pages={3739--3751},
  year={2023},
  publisher={IEEE}
}

@article{2019_manufacturing_variations,
  title={Stochastic optimal control to minimize the impact of manufacturing variations on nanomechanical systems},
  author={Ito, Y. and Funayama, K. and Hirotani, J. and Ohno, Y. and Tadokoro, Y.},
  journal={IEEE Access},
  volume={7},
  pages={171195--171205},
  year={2019},
  publisher={IEEE}
}

@article{2023_reviewer_recommendation_15,
  title={On the inherent distributional robustness of stochastic and nominal model predictive control},
  author={McAllister, R. D. and Rawlings, J. B.},
  journal={IEEE Trans. on Autom. Control},
  volume={69},
  number={2},
  pages={741--754},
  year={2023},
  publisher={IEEE}
}

@inproceedings{2021_reviewer_recommendation_9,
  title={Stochastic exponential stability of nonlinear stochastic model predictive control},
  author={McAllister, R. D. and Rawlings, J. B.},
  booktitle={60th Conf. on Decision and Control},
  pages={880--885},
  year={2021},
  organization={IEEE}
}

@article{2022_reviewer_recommendation_11,
  title={The stochastic robustness of nominal and stochastic model predictive control},
  author={McAllister, R. D. and Rawlings, J. B.},
  journal={IEEE Trans. on Autom. Control},
  volume={68},
  number={10},
  pages={5810--5822},
  year={2022},
  publisher={IEEE}
}

@article{2022_reviewer_recommendation_13,
  title={Nonlinear stochastic model predictive control: Existence, measurability, and stochastic asymptotic stability},
  author={McAllister, R. D. and Rawlings, J. B.},
  journal={IEEE Trans. on Autom. Control},
  volume={68},
  number={3},
  pages={1524--1536},
  year={2022},
  publisher={IEEE}
}

@article{2023_reviewer_recommendation_14,
  title={A suboptimal economic model predictive control algorithm for large and infrequent disturbances},
  author={McAllister, R. D. and Rawlings, J. B.},
  journal={IEEE Trans. on Autom. Control},
  volume={69},
  number={2},
  pages={1242--1248},
  year={2023},
  publisher={IEEE}
}

@article{2021_reviewer_recommendation_12,
  title={Inherent stochastic robustness of model predictive control to large and infrequent disturbances},
  author={McAllister, R. D. and Rawlings, J. B.},
  journal={IEEE Trans. on Autom. Control},
  volume={67},
  number={10},
  pages={5166--5178},
  year={2021},
  publisher={IEEE}
}

@article{2021_reviewer_recommendation_8,
  title={Robustness of model predictive control to (large) discrete disturbances},
  author={McAllister, R. D. and Rawlings, J. B.},
  journal={IFAC-PapersOnLine},
  volume={54},
  number={6},
  pages={64--69},
  year={2021},
  publisher={Elsevier}
}

@inproceedings{2022_reviewer_recommendation_10,
  title={Advances in mixed-integer model predictive control},
  author={McAllister, R. D. and Rawlings, J. B.},
  booktitle={American control conf.},
  pages={364--369},
  year={2022},
  organization={IEEE}
}

@article{2025_braatz_pc,
  title={Probabilistically Robust Uncertainty Analysis and Optimal Control of Continuous Lyophilization via Polynomial Chaos Theory},
  author={Srisuma, P. and Barbastathis, G. and Braatz, R. D.},
  journal={arXiv preprint arXiv:2502.13420},
  year={2025}
}

@article{2025_performance_nominal,
  title={Closed-loop performance optimization of model predictive control with robustness guarantees},
  author={Zuliani, R. and Balta, E. C. and Lygeros, J.},
  journal={European Journal of Control},
  pages={101319},
  year={2025},
  publisher={Elsevier}
}

@incollection{2007_robust_mpc_survey,
  title={Robust model predictive control: A survey},
  author={Bemporad, A. and Morari, M.},
  booktitle={Robustness in identification and control},
  pages={207--226},
  year={2007},
  publisher={Springer}
}

@inproceedings{2001_limitation_adaptive_robust,
  title={Fundamental limitations and differences of robust and adaptive control},
  author={Wang, L. Y and Zhang, J. F.},
  booktitle={American Control Conf.},
  volume={6},
  pages={4802--4807},
  year={2001},
  organization={IEEE}
}

@article{1987_comparison_robust_adaptive,
  title={A comparison between robust and adaptive control of uncertain systems},
  author={{\AA}str{\"o}m, K. J. and Neumann, L. and Gutman, P. O.},
  journal={IFAC Adaptive Systems in Control and Signal Processing},
  volume={20},
  number={2},
  pages={43--48},
  year={1987},
  publisher={Elsevier}
}

@book{2006_probability_rao,
  title={Probability theory with applications},
  author={Rao, M. M. and Swift, R. J.},
  year={2006},
  publisher={Springer}
}

@article{1993_stability_Meyn_Tweedie,
  title={Stability of {Markovian} processes {III}: {Foster--Lyapunov} criteria for continuous-time processes},
  author={Meyn, S. P. and Tweedie, R. L.},
  journal={Advances in Applied Probability},
  volume={25},
  number={3},
  pages={518--548},
  year={1993},
  publisher={Cambridge University Press}
}

@article{2015_mayne_robust_stochastic,
  title={Robust and stochastic {MPC}: Are we going in the right direction?},
  author={Mayne, D. Q.},
  journal={IFAC-PapersOnLine},
  volume={48},
  number={23},
  pages={1--8},
  year={2015},
  publisher={Elsevier}
}

@article{2024_F16,
	title={Model Predictive Control Design under Stochastic Parametric Uncertainties Based on Polynomial Chaos Expansions for {F-16} Aircraft},
	author={Purnawan, H. and Asfihani, T. and Kim, S. and Subchan, S.},
	journal={Journal of Robotics and Control},
	volume={5},
	number={3},
	pages={723--732},
	year={2024}
}

@article{2022_acados,
	title={acados—a modular open-source framework for fast embedded optimal control},
	author={Verschueren, R. and Frison, G. and Kouzoupis, D. and Frey, J. and Duijkeren, N. van and Zanelli, A. and Novoselnik, B. and Albin, T. and Quirynen, R. and Diehl, M.},
	journal={Mathematical Programming Computation},
	volume={14},
	number={1},
	pages={147--183},
	year={2022}
}

@inproceedings{2023_av,
	title={A stochastic nonlinear model predictive control with an uncertainty propagation horizon for autonomous vehicle motion control},
	author={Zarrouki, B. and Wang, C. and Betz, J.},
	booktitle={American Control Conf.},
    pages={5466--5473},
	year={2024}
}

@article{2024_arbitrary_power,
	title={Arbitrary polynomial chaos-based power system dynamic analysis with correlated uncertainties},
	author={Li, X. and Liu, C. and Wang, C. and Milano, F.},
	journal={International Jr. of Electrical Power \& Energy Systems},
	volume={157},
	pages={109806},
	year={2024}
}

@article{2013_separation_principle,
  title={The separation principle in stochastic control, redux},
  author={Georgiou, T. T. and Lindquist, A.},
  journal={IEEE Trans. on Autom. Control},
  volume={58},
  number={10},
  pages={2481--2494},
  year={2013}
}

@article{2014_mpc_survey_Mayne,
  title={Model predictive control: Recent developments and future promise},
  author={Mayne, D. Q.},
  journal={Automatica},
  volume={50},
  number={12},
  pages={2967--2986},
  year={2014}
}

@article{2011_inherent_robustness_conditions,
  title={Conditions under which suboptimal nonlinear {MPC} is inherently robust},
  author={Pannocchia, G. and Rawlings, J. B. and Wright, S. J.},
  journal={Systems \& Control Letters},
  volume={60},
  number={9},
  pages={747--755},
  year={2011}
}

@article{2014_inherent_robustness,
  title={Inherent robustness properties of quasi-infinite horizon nonlinear model predictive control},
  author={Yu, S. and Reble, M. and Chen, H. and Allg{\"o}wer, F.},
  journal={Automatica},
  volume={50},
  number={9},
  pages={2269--2280},
  year={2014}
}

@article{2004_nonrobustness_examples,
  title={Examples when nonlinear model predictive control is nonrobust},
  author={Grimm, G. and Messina, M. J. and Tuna, S. E. and Teel, A. R.},
  journal={Automatica},
  volume={40},
  number={10},
  pages={1729--1738},
  year={2004}
}

@article{1972_Ogura,
	title={Orthogonal functionals of the {Poisson} process},
	author={Ogura, H.},
	journal={IEEE Trans. on Information Theory},
	volume={18},
	number={4},
	pages={473--481},
	year={1972}
}

@book{2017_Rawlings_book,
	title={Model predictive control: theory, computation, and design},
	author={Rawlings, J. B. and Mayne, D. Q. and Diehl, M.},
	year={2017},
	publisher={Nob Hill Publishing Madison, WI}
}

@incollection{2018_robustness_analysis,
	title = {Robustness analysis with real parametric uncertainty},
	booktitle= {The Control Systems Handbook},
	author= {Tempo, R. and Blanchini, F.},
	chapter= {7},
	pages={7-1 -- 7-18},
	year={2018},
	publisher= {CRC Press}
}

@book{2002_nonlinear_Vidyasagar,
	title={Nonlinear Systems Analysis},
	author={Vidyasagar, M.},
	year={2002},
	publisher={SIAM}
}

@book{2012_open_problems,
	title={Open Problems in Mathematical Systems and Control Theory},
	author={Blondel, V. D. and Sontag, E. D. and Vidyasagar, M. and Willems, J. C.},
	year={2012},
	publisher={Springer Science \& Business Media}
}

@article{2020_clot,
	title={{CLOT} norm minimization for continuous hands-off control},
	author={Nagahara, M. and Chatterjee, D. and Challapalli, N. and Vidyasagar, M.},
	journal={Automatica},
	volume={113},
	year={2020}
}

@article{2023_NARX_io_model,
	title={Polynomial {NARX}-based nonlinear model predictive control of modular chemical systems},
	author={Nikolakopoulou, A. and Braatz, R. D.},
	journal={Computers \& Chemical Engineering},
	volume={177},
	year={2023}
}

@article{2010_time_dependent_pc,
	title={Time-dependent generalized polynomial chaos},
	author={Gerritsma, M. and van der Steen, J.-B. and Vos, P. and Karniadakis, G.},
	journal={Journal of Computational Physics},
	volume={229},
	number={22},
	pages={8333--8363},
	year={2010}
}

@article{2020_von_unstable,
	title={Fast stochastic model predictive control of unstable dynamical systems},
	author={von Andrian, M. and Braatz, R. D.},
	journal={IFAC-PapersOnLine},
	volume={53},
	number={2},
	pages={7262--7267},
	year={2020}
}

@article{2019_Harinath_polynomial,
	title={Model predictive control of polynomial systems},
	author={Harinath, E. and Foguth, L. C. and Paulson, J. A. and Braatz, R. D.},
	journal={Handbook of Model Predictive Control},
	pages={221--237},
	year={2019}
}

@inproceedings{2019_von_andrien_offset_free,
	title={Offset-free input-output formulations of stochastic model predictive control based on polynomial chaos theory},
	author={von Andrian, M. and Braatz, R. D.},
	booktitle={American Control Conf.},
	pages={360--365},
	year={2019}
}

@article{2013_Ikonen,
	title={Model Predictive Control and State Estimation},
	author={Ikonen, E.},
	journal={University of Oulu, Finland},
	year={2013}
}

@inproceedings{2014_fast_braatz,
	title={Fast stochastic model predictive control of high-dimensional systems},
	author={Paulson, J. A. and Mesbah, A. and Streif, S. and Findeisen, R. and Braatz, R. D.},
	booktitle={IEEE Conf. on Decision and Control},
	pages={2802--2809},
	year={2014}
}

@article{2011_konda_estimation,
	title={State uncertainty propagation in the presence of parametric uncertainty and additive white noise},
	author={Konda, U. and Singla, P. and Singh, T. and Scott, P. D.},
	journal={Journal of Dynamic Systems, Measurement, and Control},
	volume={133},
	number={5},
	year={2011}
}

@article{2010_nonlinear_estimation,
	title={Nonlinear estimation of hypersonic state trajectories in {Bayesian} framework with polynomial chaos},
	author={Dutta, P. and Bhattacharya, R.},
	journal={Journal of Guidance, Control, and Dynamics},
	volume={33},
	number={6},
	pages={1765--1778},
	year={2010}
}

@article{2010_KF_PCT,
	title={A polynomial chaos-based {Kalman} filter approach for parameter estimation of mechanical systems},
	author={Blanchard, E. D. and Sandu, A. and Sandu, C.},
	year={2010},
	journal = {Journal of Dynamic Systems, Measurement and Control},
	volume = {132},
	number = {6},
	pages = {061404-1 - 061404-18}
}

@article{2009_enKF_PCT,
	title={A generalized polynomial chaos based ensemble {Kalman} filter with high accuracy},
	author={Li, J. and Xiu, D.},
	journal={Journal of Computational Physics},
	volume={228},
	number={15},
	pages={5454--5469},
	year={2009}
}

@inproceedings{learning_MPC_Mesbah,
	title={Learning-based Stochastic Model Predictive Control with State-Dependent Uncertainty},
	author={Bonzanini, A. D. and Mesbah, A.},
	booktitle={Learning for Dynamics and Control},
	pages={571--580},
	year={2020}
}

@article{lbmpc_linear,
	title={Provably safe and robust learning-based model predictive control},
	author={Aswani, A. and Gonzalez, H. and Sastry, S. S. and Tomlin, C.},
	journal={Automatica},
	volume={49},
	number={5},
	pages={1216--1226},
	year={2013}
}

@article{Bonnaire,
	title={Intrusive generalized polynomial chaos with asynchronous time integration for the solution of the unsteady Navier–Stokes equations},
	author={P. Bonnaire and P. Pettersson and C. F. Silva},
	journal={Computers \& Fluids},
	volume={223},
	number={5},
	pages={104952},
	year={2021}
}

@article{LMPC_Borrelli,
	title={Learning model predictive control for iterative tasks. a data-driven control framework},
	author={Rosolia, U. and Borrelli, F.},
	journal={IEEE Trans. on Autom. Control},
	volume={63},
	number={7},
	pages={1883--1896},
	year={2017}
}

@article{PG20,
	title={Deep Model Predictive Control with Stability Guarantees},
	author={Mishra, P. K. and Gasparino, M. V.  and Chowdhary, G.},
	journal={IEEE Trans. on Autom. Control},
volume = {70},
issue = {8},
pages = {5460 - 5467},
	year={2025}
}

@inproceedings{2023_unmatched,
title={Unmatched uncertainty mitigation through neural network supported model predictive control},
	author={Gasparino, M. V. and Mishra, P. K. and Chowdhary, G.},
	booktitle={62nd IEEE Conf. on Decision and Control},
  pages={3555--3560},
  year={2023},
  organization={IEEE}
}

@article{survey_LMPC,
	title={Learning-based model predictive control: Toward safe learning in control},
	author={Hewing, L. and Wabersich, K. P. and Menner, M. and Zeilinger, M. N.},
	journal={Annual Review of Control, Robotics, and Autonomous Systems},
	volume={3},
	pages={269--296},
	year={2020}
}

@article{safe_learning_survey_2022,
	title={Safe learning in robotics: From learning-based control to safe reinforcement learning},
	author={Brunke, L. and Greeff, M. and Hall, A. W and Yuan, Z. and Zhou, S. and Panerati, J. and Schoellig, A. P.},
	journal={Annual Review of Control, Robotics, and Autonomous Systems},
	volume={5},
	pages={411--444},
	year={2022}
}

@article{2016_power_system,
	title={A novel method of polynomial approximation for parametric problems in power systems},
	author={Zhou, Y. and Wu, H. and Gu, C. and Song, Y.},
	journal={IEEE Trans. on Power Systems},
	volume={32},
	number={4},
	pages={3298--3307},
	year={2016}
}

@article{2007_pcm,
	title={Efficient collocational approach for parametric uncertainty analysis},
	author={Xiu, D.},
	journal={Communications in Computational Physics},
	volume={2},
	number={2},
	pages={293--309},
	year={2007}
}

@book{1984_galerkin,
	title={Computational Galerkin Methods},
	author={Fletcher, C. A. J.},
	year={1984},
	publisher={Springer}
}

@article{1988_biorthogonal,
	title={On the theory of biorthogonal polynomials},
	author={Iserles, A. and N{\o}rsett, S. P.},
	journal={Trans. of the American Mathematical Society},
	volume={306},
	number={2},
	pages={455--474},
	year={1988}
}

@article{1982_stieltjes_procedure,
	title={On generating orthogonal polynomials},
	author={Gautschi, W.},
	journal={SIAM Journal on Scientific and Statistical Computing},
	volume={3},
	number={3},
	pages={289--317},
	year={1982}
}

@article{2003_xiu_kle,
	title={Modeling uncertainty in flow simulations via generalized polynomial chaos},
	author={Xiu, D. and Karniadakis, G. E.},
	journal={Journal of Computational Physics},
	volume={187},
	number={1},
	pages={137--167},
	year={2003}
}

@article{2016_regression,
	title={Nonadaptive quasi-optimal points selection for least squares linear regression},
	author={Shin, Y. and Xiu, D.},
	journal={SIAM Journal on Scientific Computing},
	volume={38},
	number={1},
	pages={A385--A411},
	year={2016}
}

@article{2010_pseudospectral,
	title={Spectral methods for parameterized matrix equations},
	author={Constantine, P. G. and Gleich, D. F. and Iaccarino, G.},
	journal={SIAM Journal on Matrix Analysis and Applications},
	volume={31},
	number={5},
	pages={2681--2699},
	year={2010}
}

@article{2021_discounted_constraint,
	title={Stochastic {MPC} with dynamic feedback gain selection and discounted probabilistic constraints},
	author={Yan, S. and Goulart, P. J. and Cannon, M.},
	journal={IEEE Trans. on Autom. Control},
	volume={67},
	number={11},
	pages={5885--5899},
	year={2021}
}

@article{2014_korda_average_constraint_violation,
	title={Stochastic {MPC} framework for controlling the average constraint violation},
	author={Korda, M. and Gondhalekar, R. and Oldewurtel, F. and Jones, C. N.},
	journal={IEEE Trans. on Autom. Control},
	volume={59},
	number={7},
	pages={1706--1721},
	year={2014}
}

@inproceedings{2021_recursive_feasibility,
	title={Recursive feasibility of stochastic model predictive control with mission-wide probabilistic constraints},
	author={Wang, K. and Gros, S.},
	booktitle={IEEE Conf. on Decision and Control},
	pages={2312--2317},
	year={2021}
}

@article{2021_constraint_violation,
	title={Minimization of constraint violation probability in model predictive control},
	author={Br{\"u}digam, T. and Ga{\ss}mann, V. and Wollherr, D. and Leibold, M.},
	journal={Int. Journal of Robust and Nonlinear Control},
	volume={31},
	number={14},
	pages={6740--6772},
	year={2021}
}

@article{2006_chance_constraint_approximation,
	title={Tractable approximations to robust conic optimization problems},
	author={Bertsimas, D. and Sim, M.},
	journal={Mathematical Programming},
	volume={107},
	number={1-2},
	pages={5--36},
	year={2006}
}

@article{2009_primbs_recusrsive_feasibility,
	title={Stochastic receding horizon control of constrained linear systems with state and control multiplicative noise},
	author={Primbs, J. A. and Sung, C. H.},
	journal={IEEE Trans. on Autom. Control},
	volume={54},
	number={2},
	pages={221--230},
	year={2009}
}

@inproceedings{ref:amin-10,
	title={Mean-square boundedness of stochastic networked control systems with bounded control inputs},
	author={Chatterjee, D. and Amin, S. and Hokayem, P. and Lygeros, J. and Sastry, S. S.},
	booktitle={Conf. on Decision and Control},
	pages={4759--4764},
	year={2010}
}

@inproceedings{ref:HokChaRamChaLyg-10,
	AUTHOR = {P. Hokayem and D. Chatterjee and F. Ramponi and G. Chaloulos and J. Lygeros},
	TITLE = {Stable stochastic receding horizon control of linear systems with bounded control inputs},
	BOOKTITLE = {Int. Symposium on Mathematical Theory of Networks and Systems, Budapest, Hungary},
	YEAR = {2010},
	PAGES = {31--36},
}

@article{goulart-06,
	title={Optimization over state feedback policies for robust control with constraints},
	author={Goulart, P. J. and Kerrigan, E. C. and Maciejowski, J. M.},
	journal={Automatica},
	volume={42},
	number={4},
	pages={523--533},
	year={2006}
}

@book{kumar1986stochastic,
	title={Stochastic Systems: Estimation, Identification and Adaptive Control},
	author={Kumar, P. R. and Varaiya, P.},
	year={1986},
	publisher={Information and System Sciences Series. Prentice Hall, Upper Saddle River, New Jersey}
}

@article{2015_paulson_ti,
	title={Nonlinear model predictive control of systems with probabilistic time-invariant uncertainties},
	author={Paulson, J. A. and Harinath, E. and Foguth, L. C. and Braatz, R. D.},
	journal={IFAC-PapersOnLine},
	volume={48},
	number={23},
	pages={16--25},
	year={2015}
}

@article{2014_chatterjee_lygeros,
	title={On stability and performance of stochastic predictive control techniques},
	author={Chatterjee, D. and Lygeros, J.},
	journal={IEEE Trans. on Autom. Control},
	volume={60},
	number={2},
	pages={509--514},
	year={2014}
}

@article{2018_stabilizing_conditions,
	title={Stabilizing conditions for model predictive control},
	author={Mayne, D. Q. and Falugi, P.},
	journal={Int. Journal of Robust and Nonlinear Control},
	volume={29},
	number={4},
	pages={894--903},
	year={2018}
}

@article{2017_mesbah,
	title={Stochastic model predictive control with joint chance constraints},
	author={Paulson, J. A. and Buehler, E. A. and Braatz, R. D. and Mesbah, A.},
	journal={Int. Journal of Control},
	volume={93},
	number={1},
	pages={126--139},
	year={2020}
}

@inproceedings{bosgra2003,
	title={A full solution to the constrained stochastic closed-loop {MPC} problem via state and innovations feedback and its receding horizon implementation},
	author={van Hessem, D. H. and Bosgra, O. H.},
	booktitle={IEEE Conf. on Decision and Control},
	pages={929--934},
	year={2003}
}

@inproceedings{2019_evolving_disturbance,
	title={Input hard constrained optimal covariance steering},
	author={Okamoto, K. and Tsiotras, P.},
	booktitle={IEEE Conf. on Decision and Control},
	pages={3497--3502},
	year={2019}
}

@article{2021_mishra_automatica,
	title={Reference tracking stochastic model predictive control over unreliable channels and bounded control actions},
	author={Mishra, P. K. and Diwale, S. S. and Jones, C. N. and Chatterjee, D.},
	journal={Automatica},
	volume={127},
	year={2021}
}

@article{2020_mishra_automatica,
	title={Stochastic predictive control under intermittent observations and unreliable actions},
	author={Mishra, P. K. and Chatterjee, D. and Quevedo, D. E.},
	journal={Automatica},
	volume={118},
	year={2020}
}

@article{prabhatNOLCOS2016, 
	author={Mishra, P. K. and Quevedo, D. E. and Chatterjee, D.},	 
	title={Dropout feedback parametrized policies for stochastic predictive controller}, 
	journal={IFAC-PapersOnLine}, 
	volume={49},
	number={18},
	pages={59--64},
	year={2016}
}

@article{ref:Hokayem-12,
	author = {Hokayem, P. and Cinquemani, E. and Chatterjee, D. and Ramponi, F. and Lygeros, J.},
	journal = {Automatica},
	number = {1},
	pages = {77--88},
	title = {{Stochastic receding horizon control with output feedback and bounded controls}},
	volume = {48},
	year = {2012}
}

@inproceedings{hokayem2009stochastic,
	title={On stochastic receding horizon control with bounded control inputs},
	author={Hokayem, P. and Chatterjee, D. and Lygeros, J.},
	booktitle={IEEE Conf. on Decision and Control, held jointly with the Chinese Control Conference},
	pages={6359--6364},
	year={2009}
}

@article{garstka1974decision,
	title={On decision rules in stochastic programming},
	author={Garstka, S. J. and Wets, R. J.-B.},
	journal={Mathematical Programming},
	volume={7},
	number={1},
	pages={117--143},
	year={1974}
}

@inproceedings{2009_convex_constraints,
	title={Convex chance constrained predictive control without sampling},
	author={Blackmore, L. and Ono, M.},
	booktitle={AIAA Guidance, Navigation, and Control Conf.},
	pages={5876},
	year={2009}
}

@article{2005_sensitivity_sobol,
	title={Global sensitivity indices for nonlinear mathematical models, Review},
	author={Sobol, I. and Kucherenko, S.},
	journal={Wilmott Mag},
	volume={1},
	pages={56--61},
	year={2005}
}

@book{1994_kernel_smoothing,
	title={Kernel Smoothing},
	author={Wand, M. P. and Jones, M. C.},
	year={1994},
	publisher={CRC Press}
}

@inproceedings{2013_sudret_sparse,
	title={Sparse polynomial chaos expansions of vector-valued response quantities},
	author={Blatman, G. and Sudret, B.},
	booktitle={Int. Conf. Struct. Safety and Reliability, Newyork, USA},
	year={2013}
}

@inproceedings{2011_sudret_pca,
	title={Principal component analysis and Least Angle Regression in spectral stochastic finite element analysis},
	author={Blatman, G. and Sudret, B.},
	booktitle={Int. Conf. on Applications of Stat. and Prob. in Civil Engineering, Zurich, Switzerland},
	year={2011}
}

@article{2004_lar,
	title={Least angle regression},
	author={Efron, B. and Hastie, T. and Johnstone, I. and Tibshirani, R.},
	journal={Annals of Statistics},
	volume={32},
	pages={407--499},
	year={2004}
}

@article{2010_blatman_Sudret,
	title={An adaptive algorithm to build up sparse polynomial chaos expansions for stochastic finite element analysis},
	author={Blatman, G. and Sudret, B.},
	journal={Probabilistic Engineering Mechanics},
	volume={25},
	number={2},
	pages={183--197},
	year={2010}
}

@article{2006_sudret,
	title={Global sensitivity analysis using polynomial chaos expansions},
	author={Sudret, B.},
	journal={Reliability Engineering \& System Safety},
	volume={93},
	number={7},
	pages={964--979},
	year={2008}
}

@book{1996_reliability_methods,
	title={Structural Reliability Methods},
	author={Ditlevsen, O. and Madsen, H. O.},
	volume={178},
	year={1996},
	publisher={John Wiley \& Sons Ltd, England}
}

@article{2014_PCE_Sudret,
	title={Polynomial chaos expansions and stochastic finite element methods},
	author={Sudret, B.},
	journal={Risk and Reliability in Geotechnical Engineering},
	pages={265--300},
	year={2014}
}

@Article{casadi,
	Author = {Andersson, J. A. E. and Gillis, J. and Horn, G. 
	and Rawlings, J. B.  and Diehl, M.},
	Title = {{CasADi} -- {A} software framework for nonlinear optimization
	and optimal control},
	Journal = {Mathematical Programming Computation},
	volume={11},
	number={1},
	pages={1--36},
	year={2019}
}

@article{2004_oed,
	title={Numerical methods for optimal control problems in design of robust optimal experiments for nonlinear dynamic processes},
	author={K{\"o}rkel, S. and Kostina, E. and Bock, H. G. and Schl{\"o}der, J. P.},
	journal={Optimization Methods and Software},
	volume={19},
	number={3-4},
	pages={327--338},
	year={2004}
}

@article{2023_PCE_GP,
	title={Recent advances in uncertainty quantification methods for engineering problems},
	author={Kumar, D. and Ahmed, F. and Usman, S. and Alajo, A. and Alam, S. B.},
	journal={AI Assurance},
	pages={453--472},
	year={2023}
}

@incollection{2014_uqlab,
	title={UQLab: A framework for uncertainty quantification in Matlab},
	author={Marelli, S. and Sudret, B.},
	booktitle={Vulnerability, Uncertainty, and Risk: Quantification, Mitigation, and Management},
	pages={2554--2563},
	year={2014}
}

@article{2014_pce_correlated,
	title={Polynomial chaos expansion for general multivariate distributions with correlated variables},
	author={Navarro, M. and Witteveen, J. and Blom, J.},
	journal={arXiv preprint arXiv:1406.5483},
	year={2014}
}

@article{2012_convergence_gPC,
	title={On the convergence of generalized polynomial chaos expansions},
	author={Ernst, O. G. and Mugler, A. and Starkloff, H. and Ullmann, E.},
	journal={{ESAIM}: Mathematical Modelling and Numerical Analysis},
	volume={46},
	number={2},
	pages={317--339},
	year={2012}
}

@book{2009_robust_optimization,
	title={Robust Optimization},
	author={Ben-Tal, A. and El Ghaoui, L. and Nemirovski, A.},
	volume={28},
	year={2009},
	publisher={Princeton University Press}
}

@article{2009_saa,
	title={Sample average approximation method for chance constrained programming: theory and applications},
	author={Pagnoncelli, B. K. and Ahmed, S. and Shapiro, A.},
	journal={Journal of Optimization Theory and Applications},
	volume={142},
	number={2},
	pages={399--416},
	year={2009}
}

@article{2007_convex_approximation,
	title={Convex approximations of chance constrained programs},
	author={Nemirovski, A. and Shapiro, A.},
	journal={SIAM Journal on Optimization},
	volume={17},
	number={4},
	pages={969--996},
	year={2007}
}

@article{2000_discretized_pdf,
	title={Concavity and efficient points of discrete distributions in probabilistic programming},
	author={Dentcheva, D. and Pr{\'e}kopa, A. and Ruszczynski, A.},
	journal={Mathematical Programming},
	volume={89},
	pages={55--77},
	year={2000}
}

@book{2015_MPC_book,
	title={Model Predictive Control: Classical, Robust and Stochastic},
	author={Kouvaritakis, B. and Cannon, M.},
	year={2015},
	publisher={Springer International Publishing}
}

@article{2016_constraint_tightening,
	title={Constraint-tightening and stability in stochastic model predictive control},
	author={Lorenzen, M. and Dabbene, F. and Tempo, R. and Allg{\"o}wer, F.},
	journal={IEEE Trans. on Autom. Control},
	volume={62},
	number={7},
	pages={3165--3177},
	year={2016}
}

@article{2012_oldewurtel,
	title={Use of model predictive control and weather forecasts for energy efficient building climate control},
	author={Oldewurtel, F. and Parisio, A. and Jones, C. N. and Gyalistras, D. and Gwerder, M. and Stauch, V. and Lehmann, B. and Morari, M.},
	journal={Energy and Buildings},
	volume={45},
	pages={15--27},
	year={2012}
}

@article{2014_scenario,
	title={The scenario approach for stochastic model predictive control with bounds on closed-loop constraint violations},
	author={Schildbach, G. and Fagiano, L. and Frei, C. and Morari, M.},
	journal={Automatica},
	volume={50},
	number={12},
	pages={3009--3018},
	year={2014}
}

@article{2013_scenario,
	title={Stochastic model predictive control of {LPV} systems via scenario optimization},
	author={Calafiore, G. C. and Fagiano, L.},
	journal={Automatica},
	volume={49},
	number={6},
	pages={1861--1866},
	year={2013}
}

@article{2009_sequential_mc,
	title={Sequential {Monte Carlo} for model predictive control},
	author={Kantas, N. and Maciejowski, J. and Visintini, A. L.},
	journal={Nonlinear Model Predictive Control: Towards New Challenging Applications},
	pages={263--273},
	year={2009}
}

@article{2006_mc,
	title={{Monte Carlo} optimization for conflict resolution in air traffic control},
	author={Visintini, A. L. and Glover, W. and Lygeros, J. and Maciejowski, J.},
	journal={IEEE Trans. on Intelligent Transportation Systems},
	volume={7},
	number={4},
	pages={470--482},
	year={2006}
}

@article{2004_unscented_filtering,
	title={Unscented filtering and nonlinear estimation},
	author={Julier, S. J. and Uhlmann, J. K.},
	journal={Proceedings of the IEEE},
	volume={92},
	number={3},
	pages={401--422},
	year={2004}
}

@article{2000_linearized_propagation,
	title={Decreasing the sensitivity of open-loop optimal solutions in decision making under uncertainty},
	author={Darlington, J. and Pantelides, C. C. and Rustem, B. and Tanyi, B. A.},
	journal={European Journal of Operational Research},
	volume={121},
	number={2},
	pages={343--362},
	year={2000}
}

@book{2018_uq,
	title={Uncertainty Quantification and Predictive Computational Science},
	author={McClarren, R. G.},
	year={2018},
	publisher={Springer}
}

@book{2015_sullivan,
	title={Introduction to Uncertainty Quantification},
	author={Sullivan, T. J.},
	volume={63},
	year={2015},
	publisher={Springer}
}

@book{2010_xiu,
	title={Numerical Methods for Stochastic Computations: A Spectral Method Approach},
	author={Xiu, D.},
	year={2010},
	publisher={Princeton University Press}
}

@article{1985_askey_scheme,
	title={Some basic hypergeometric orthogonal polynomials that generalize {Jacobi} polynomials},
	author={Askey, R. and Wilson, J. A.},
	volume={319},
	year={1985},
	journal={Memoirs of the American Mathematical Soc.}
}

@article{1952_rosenblatt_transformation,
	title={Remarks on a multivariate transformation},
	author={Rosenblatt, M.},
	journal={The Annals of Mathematical Statistics},
	volume={23},
	number={3},
	pages={470--472},
	year={1952}
}

@article{2020_polympc,
  title={{PolyMPC}: An efficient and extensible tool for real-time nonlinear model predictive tracking and path following for fast mechatronic systems},
  author={Listov, P. and Jones, C.},
  journal={Optimal Control Applications and Methods},
  volume={41},
  number={2},
  pages={709--727},
  year={2020}
}

@article{2020_mars_life_support,
  title={Fast probabilistic uncertainty quantification and sensitivity analysis of a mars life support system model},
  author={Makrygiorgos, G. and Gupta, S. S. and Menezes, A. A. and Mesbah, A.},
  journal={IFAC-PapersOnLine},
  volume={53},
  number={2},
  pages={7268--7273},
  year={2020}
}

@article{2018_pharmaceutical,
  title={Fast stochastic model predictive control of end-to-end continuous pharmaceutical manufacturing},
  author={Paulson, J. A. and Streif, S. and Findeisen, R. and Braatz, R. D. and Mesbah, A.},
  journal={Computer Aided Chemical Engineering},
  volume={41},
  pages={353--378},
  year={2018}
}

@inproceedings{2018_apc,
  title={Shaping the closed-loop behavior of nonlinear systems under probabilistic uncertainty using arbitrary polynomial chaos},
  author={Paulson, J. A. and Mesbah, A.},
  booktitle={IEEE Conf. on Decision and Control},
  pages={6307--6313},
  year={2018}
}

@article{2019_output_Imsland,
  title={Output feedback stochastic nonlinear model predictive control for batch processes},
  author={Bradford, E. and Imsland, L.},
  journal={Computers \& Chemical Engineering},
  volume={126},
  pages={434--450},
  year={2019}
}

@article{2019_nonsmooth_PCT,
  title={Fast uncertainty quantification for dynamic flux balance analysis using non-smooth polynomial chaos expansions},
  author={Paulson, J. A. and Martin-Casas, M. and Mesbah, A.},
  journal={PLoS Computational Biology},
  volume={15},
  number={8},
  year={2019}
}

@article{2020_surrogate_modeling,
  title={Surrogate modeling for fast uncertainty quantification: Application to {2D} population balance models},
  author={Makrygiorgos, G. and Maggioni, G. M. and Mesbah, A.},
  journal={Computers \& Chemical Engineering},
  volume={138},
  pages={106814},
  year={2020}
}

@article{2013_PCE_Bayesian,
  title={Polynomial-chaos-based {Bayesian} approach for state and parameter estimations},
  author={Madankan, R. and Singla, P. and Singh, T. and Scott, P. D.},
  journal={Journal of Guidance, Control, and Dynamics},
  volume={36},
  number={4},
  pages={1058--1074},
  year={2013}
}

@article{2015_chaospy,
  title={Chaospy: An open source tool for designing methods of uncertainty quantification},
  author={Feinberg, J. and Langtangen, H. P.},
  journal={Journal of Computational Science},
  volume={11},
  pages={46--57},
  year={2015}
}

@article{comparison_PCE_sensitivity,
  title={Optimal experiment design under parametric uncertainty: a comparison of a sensitivities based approach versus a polynomial chaos based stochastic approach},
  author={Nimmegeers, P. and Bhonsale, S. and Telen, D. and Van Impe, J.},
  journal={Chemical Engineering Science},
  volume={221},
  year={2020}
}

@article{2017_Mesbah_IJRNC,
  title={An efficient method for stochastic optimal control with joint chance constraints for nonlinear systems},
  author={Paulson, J. A. and Mesbah, A.},
  journal={Int. Journal of Robust and Nonlinear Control},
  volume={29},
  number={15},
  pages={5017--5037},
  year={2017}
}

@article{uncertainty_propagation_comparison_2016,
  title={Dynamic optimization of biological networks under parametric uncertainty},
  author={Nimmegeers, P. and Telen, D. and Logist, F. and Impe, J. V.},
  journal={BMC Systems Biology},
  volume={10},
  pages={1--20},
  year={2016}
}

@article{swarms_2014,
  title={Model predictive control of swarms of spacecraft using sequential convex programming},
  author={Morgan, D. and Chung, S. and Hadaegh, F. Y.},
  journal={Journal of Guidance, Control, and Dynamics},
  volume={37},
  number={6},
  pages={1725--1740},
  year={2014}
}

@article{motion_planning_2023,
  title={Trajectory Optimization of Chance-Constrained Nonlinear Stochastic Systems for Motion Planning Under Uncertainty},
  author={Nakka, Y. K. and Chung, S.},
  journal={IEEE Trans. on Robotics},
  volume ={39}, 
  number={1},
  pages={203-222},
  year={2023}
}

@article{2016_review_SMPC,
  title={Stochastic model predictive control: An overview and perspectives for future research},
  author={Mesbah, A.},
  journal={IEEE Control Systems Magazine},
  volume={36},
  number={6},
  pages={30--44},
  year={2016}
}

@article{2023_survey_PCE_behavoral,
  title={Behavioral theory for stochastic systems: A data-driven journey from {Willems} to {Wiener} and back again},
  author={Faulwasser, T. and Ou, R. and Pan, G. and Schmitz, P. and Worthmann, K.},
  journal={Annual Reviews in Control},
  year={2023}
}

@inproceedings{2022_pseudo_linear,
  title={A State-Dependent {Riccati} Equation-Based Robust Control Approach for Nonlinear Systems with Parametric Uncertainties},
  author={Bhusal, R. and Bhattacharjee, D. and Subbarao, K.},
  booktitle={American Control Conf.},
  pages={1108--1113},
  year={2022}
}

@article{multiple_shooting_Timm_2022,
  title={Data-driven multiple shooting for stochastic optimal control},
  author={Ou, R. and Pan, G. and Faulwasser, T.},
  journal={IEEE Control Systems Letters},
  volume={7},
  pages={313--318},
  year={2022}
}

@inproceedings{input_design_2016,
  title={Stochastic predictive control with adaptive model maintenance},
  author={Bavdekar, V. A. and Ehlinger, V. and Gidon, D. and Mesbah, A.},
  booktitle={IEEE Conf. on Decision and Control},
  pages={2745--2750},
  year={2016}
}

@article{coupled_constraint_2015,
  title={Distributed model predictive control of linear systems with stochastic parametric uncertainties and coupled probabilistic constraints},
  author={Dai, L. and Xia, Y. and Gao, Y.},
  journal={SIAM Journal on Control and Optimization},
  volume={53},
  number={6},
  pages={3411--3431},
  year={2015}
}

@article{explicit_backoff_2018,
  title={Nonlinear model predictive control with explicit backoffs for stochastic systems under arbitrary uncertainty},
  author={Paulson, J. A. and Mesbah, A.},
  journal={IFAC-PapersOnLine},
  volume={51},
  number={20},
  pages={523--534},
  year={2018}
}

@inproceedings{output_feedback_2016,
  title={Output feedback model predictive control with probabilistic uncertainties for linear systems},
  author={M{\"u}hlpfordt, T. and Paulson, J. A. and Braatz, R. D. and Findeisen, R.},
  booktitle={American Control Conf.},
  pages={2035--2040},
  year={2016}
}

@inproceedings{error_bound_2017,
  title={Polynomial chaos explicit solution of the optimal control problem in model predictive control},
  author={Lefebvre, T. and De Belie, F. and Crevecoeur, G.},
  booktitle={International Conf. on Advanced Intelligent Mechatronics},
  pages={1762--1767},
  year={2017}
}

@article{truncation_error_2018,
  title={Comments on truncation errors for polynomial chaos expansions},
  author={M{\"u}hlpfordt, T. and Findeisen, R. and Hagenmeyer, V. and Faulwasser, T.},
  journal={IEEE Control Systems Letters},
  volume={2},
  number={1},
  pages={169--174},
  year={2017}
}

@article{chance_constraints_2006,
  title={On distributionally robust chance-constrained linear programs},
  author={Calafiore, G. C. and Ghaoui, L. E.},
  journal={Journal of Optimization Theory and Applications},
  volume={130},
  pages={1--22},
  year={2006}
}

@inproceedings{policy_2008,
  title={A tractable approximation of chance constrained stochastic {MPC} based on affine disturbance feedback},
  author={Oldewurtel, F. and Jones, C. N. and Morari, M.},
  booktitle={IEEE Conf. on Decision and Control},
  pages={4731--4736},
  year={2008}
}

@article{stochastic_tube_2010,
  title={Stochastic tubes in model predictive control with probabilistic constraints},
  author={Cannon, M. and Kouvaritakis, B. and Rakovi{\'c}, S. V. and Cheng, Q.},
  journal={IEEE Trans. on Autom. Control},
  volume={56},
  number={1},
  pages={194--200},
  year={2010}
}

@inproceedings{hewing_2018,
  title={Stochastic model predictive control for linear systems using probabilistic reachable sets},
  author={Hewing, L. and Zeilinger, M. N.},
  booktitle={IEEE Conf. on Decision and Control},
  pages={5182--5188},
  year={2018}
}

@article{uncertainpy_2018,
  title={Uncertainpy: a python toolbox for uncertainty quantification and sensitivity analysis in computational neuroscience},
  author={Tenn{\o}e, S. and Halnes, G. and Einevoll, G. T.},
  journal={Frontiers in Neuroinformatics},
  volume={12},
  pages={49},
  year={2018}
}

@ARTICLE{PolyChaos_2020,
       author = {{M{\"u}hlpfordt}, T. and {Zahn}, F. and {Hagenmeyer}, V. and {Faulwasser}, T.},
        title = "{PolyChaos.jl -- A Julia Package for Polynomial Chaos in Systems and Control}",
      journal={IFAC-PapersOnLine},
  volume={53},
  number={2},
  pages={7210--7216},
  year={2020},
  publisher={Elsevier}
}

@article{pocet_2020,
  title={Pocet: a polynomial chaos expansion toolbox for matlab},
  author={Petzke, F. and Mesbah, A. and Streif, S.},
  journal={IFAC-PapersOnLine},
  volume={53},
  number={2},
  pages={7256--7261},
  year={2020}
}

@article{pygpc_2020,
  title={{Pygpc}: A sensitivity and uncertainty analysis toolbox for {Python}},
  author={Weise, K. and Po{\ss}ner, L. and M{\"u}ller, E. and Gast, R. and Kn{\"o}sche, T. R.},
  journal={SoftwareX},
  volume={11},
  pages={100450},
  year={2020}
}

@article{2005_pcm,
  title={High-order collocation methods for differential equations with random inputs},
  author={Xiu, D. and Hesthaven, J. S.},
  journal={SIAM Journal on Scientific Computing},
  volume={27},
  number={3},
  pages={1118--1139},
  year={2005}
}

@article{pseudo_spectral_2013,
  title={Adaptive {Smolyak} pseudospectral approximations},
  author={Conrad, P. R. and Marzouk, Y. M.},
  journal={SIAM Journal on Scientific Computing},
  volume={35},
  number={6},
  pages={A2643--A2670},
  year={2013}
}

@article{adaptive_sparse_2011,
  title={Adaptive sparse polynomial chaos expansion based on least angle regression},
  author={Blatman, G. and Sudret, B.},
  journal={Journal of Computational Physics},
  volume={230},
  number={6},
  pages={2345--2367},
  year={2011}
}

@article{anisotropic_2008,
  title={An anisotropic sparse grid stochastic collocation method for partial differential equations with random input data},
  author={Nobile, F. and Tempone, R. and Webster, C. G.},
  journal={SIAM Journal on Numerical Analysis},
  volume={46},
  number={5},
  pages={2411--2442},
  year={2008}
}

@article{review_UQ_Fluid_2009,
  title={Uncertainty quantification and polynomial chaos techniques in computational fluid dynamics},
  author={Najm, H. N.},
  journal={Annual Review of Fluid Mechanics},
  volume={41},
  pages={35--52},
  year={2009}
}

@article{review_IC_PCE_2018,
  title={Review of polynomial chaos-based methods for uncertainty quantification in modern integrated circuits},
  author={Kaintura, A. and Dhaene, T. and Spina, D.},
  journal={Electronics},
  volume={7},
  number={3},  
  year={2018}
}

@article{power_system_2004,
  title={Evaluation of uncertainty in dynamic simulations of power system models: The probabilistic collocation method},
  author={Hockenberry, J. R. and Lesieutre, B. C.},
  journal={IEEE Trans. on Power Systems},
  volume={19},
  number={3},
  pages={1483--1491},
  year={2004}
}

@inproceedings{buehler_2016_complete_pdf,
  title={Lyapunov-based stochastic nonlinear model predictive control: Shaping the state probability distribution functions},
  author={Buehler, E. A. and Paulson, J. A. and Mesbah, A.},
  booktitle={American Control Conf.},
  pages={5389--5394},
  year={2016}
}

@inproceedings{bavdekar_2016_histogram,
  title={A polynomial chaos-based nonlinear {Bayesian} approach for estimating state and parameter probability distribution functions},
  author={Bavdekar, V. A. and Mesbah, A.},
  booktitle={American Control Conf.},
  pages={2047--2052},
  year={2016}
}

@article{mesbah_2014_diagnosis,
  title={Active fault diagnosis for nonlinear systems with probabilistic uncertainties},
  author={Mesbah, A. and Streif, S. and Findeisen, R. and Braatz, R. D.},
  journal={IFAC Proceedings Volumes},
  volume={47},
  number={3},
  pages={7079--7084},
  year={2014}
}

@inproceedings{mesbah_2014_snmpc,
  title={Stochastic nonlinear model predictive control with probabilistic constraints},
  author={Mesbah, A. and Streif, S. and Findeisen, R. and Braatz, R. D.},
  booktitle={American Control Conf.},
  pages={2413--2419},
  year={2014}
}

@article{kohler_2022,
  title={Recursively feasible stochastic predictive control using an interpolating initial state constraint},
  author={K{\"o}hler, J. and Zeilinger, M. N.},
  journal={IEEE Control Systems Letters},
  volume={6},
  pages={2743--2748},
  year={2022}
}

@article{calafiore_2012,
  title={Robust model predictive control via scenario optimization},
  author={Calafiore, G. C. and Fagiano, L.},
  journal={IEEE Trans. on Autom. Control},
  volume={58},
  number={1},
  pages={219--224},
  year={2012}
}

@article{scenario_2006,
  title={The scenario approach to robust control design},
  author={Calafiore, G. C. and Campi, M. C.},
  journal={IEEE Trans. on Autom. control},
  volume={51},
  number={5},
  pages={742--753},
  year={2006}
}

@article{scenario_2009,
  title={The scenario approach for systems and control design},
  author={Campi, M. C. and Garatti, S. and Prandini, M.},
  journal={Annual Reviews in Control},
  volume={33},
  number={2},
  pages={149--157},
  year={2009}
}

@article{blackmore_2010_particle_chance,
  title={A probabilistic particle-control approximation of chance-constrained stochastic predictive control},
  author={Blackmore, L. and Ono, M. and Bektassov, A. and Williams, B. C.},
  journal={IEEE Trans. on Robotics},
  volume={26},
  number={3},
  pages={502--517},
  year={2010}
}

@phdthesis{batina_2004_monte-carlo, 
title={Model predictive control for stochastic systems by randomized algorithms}, 
school={Technische Universiteit Eindhoven, Eindhoven, Netherlands}, 
author={Batina, I.}, 
year={2004}
}

@article{fisher_2009_lqr,
  title={Linear quadratic regulation of systems with stochastic parameter uncertainties},
  author={Fisher, J. and Bhattacharya, R.},
  journal={Automatica},
  volume={45},
  number={12},
  pages={2831--2841},
  year={2009}
}

@article{wiener_1938_homogeneous,
  title={The homogeneous chaos},
  author={Wiener, N.},
  journal={American Journal of Mathematics},
  volume={60},
  number={4},
  pages={897--936},
  year={1938}
}

@book{stochastic_finite_elements_2003,
  title={Stochastic Finite Elements: A Spectral Approach},
  author={Ghanem, R. G. and Spanos, P. D.},
  year={1991},
  publisher={Courier Corporation}
}

@inproceedings{paulson_stability_2015,
  title={Stability for receding-horizon stochastic model predictive control},
  author={Paulson, J. A. and Streif, S. and Mesbah, A.},
  booktitle={American Control Conf.},
  pages={937--943},
  year={2015}
}

@article{PCE_NN_2022,
  title={Model order reduction for real-time hybrid simulation: Comparing polynomial chaos expansion and neural network methods},
  author={Tsokanas, N. and Simpson, T. and Pastorino, R. and Chatzi, E. and Stojadinovi{\'c}, B.},
  journal={Mechanism and Machine Theory},
  volume={178},
  year={2022}
}

@article{sparse_grid_collocation_2022,
  title={Global self-optimizing control with active-set changes: A polynomial chaos approach},
  author={Ye, L. and Cao, Y. and Yang, S.},
  journal={Computers \& Chemical Engineering},
  volume={159},
  year={2022}
}

@phdthesis{winokur_2015,
  title={Adaptive sparse grid approaches to polynomial chaos expansions for uncertainty quantification},
  author={Winokur, J. G.},
  year={2015},
  school={Duke University, USA}
}

@article{Autonomous_navigation_Jones_2021,
  title={Stochastic Optimal Control for Autonomous Driving Applications via Polynomial Chaos Expansions},
  author={Listov, P. and Schwarz, J. and Jones, C.},
  journal={Optimal Control Applications and Methods},
  volume={45},
  number={1},
  pages={3--28},
  year={2024},
  publisher={Wiley Online Library}
}

@article{Imsland_Gauss_Hermite_2019,
  title={Stochastic nonlinear model predictive control of a batch fermentation process},
  author={Bradford, E. and Imsland, L.},
  journal={Computer Aided Chemical Engineering},
  volume={46},
  pages={1237--1242},
  year={2019}
}

@article{Imsland_filter_2019,
  title={Economic stochastic nonlinear model predictive control of a semi-batch polymerization reaction},
  author={Bradford, E. and Reble, M. and Bouaswaig, A. and Imsland, L.},
  journal={IFAC-PapersOnLine},
  volume={52},
  number={1},
  pages={667--672},
  year={2019}
}

@article{Chebyshev_inequality_1960,
  title={Multivariate {Chebyshev} inequalities},
  author={Marshall, A. W. and Olkin, I.},
  journal={The Annals of Mathematical Statistics},
  pages={1001--1014},
  year={1960}
}

@article{mpc_manufacturing_2021,
  title={Model predictive control in pharmaceutical continuous manufacturing: A review from a user’s perspective},
  author={Jelsch, M. and Roggo, Y. and Kleinebudde, P. and Krumme, M.},
  journal={European Journal of Pharmaceutics and Biopharmaceutics},
  volume={159},
  pages={137--142},
  year={2021}
}

@article{chance_constraint_sigmoid_2019,
  title={Rare Event Chance-Constrained Optimal Control Using Polynomial Chaos and Subset Simulation},
  author={Piprek, P. and Gros, S. and Holzapfel, F.},
  journal={Processes},
  volume={7},
  number={4},
  pages={185},
  year={2019}
}

@article{nonlinear_surrogate_PCE,
  title={Surrogate Modeling \& Optimization of a Nonlinear Batch Reactor by Polynomial Chaos Expansion},
  author={Tasnim, N. and Momtaz, M. and Sanzida, N.},
  journal={Chemical Engineering Research Bulletin},
  pages={121--126},
  year={2020}
}

@article{PCE_reduced_variables_2019,
  title={An iterative polynomial chaos approach toward stochastic elastostatic structural analysis with {non-Gaussian} randomness},
  author={Nath, K. and Dutta, A. and Hazra, B.},
  journal={Int. Journal for Numerical Methods in Engineering},
  volume={119},
  number={11},
  pages={1126--1160},
  year={2019}
}

@article{Sparse_PCE_2022,
  title={A fully {Bayesian} sparse polynomial chaos expansion approach with joint priors on the coefficients and global selection of terms},
  author={B{\"u}rkner, P. and Kr{\"o}ker, I. and Oladyshkin, S. and Nowak, W.},
  journal={Journal of Computational Physics},
  volume={488},
  pages={112210},
  year={2023},
  publisher={Elsevier}
}

@inproceedings{time_varying_2022,
  title={Polynomial Chaos Approximation of the Quadratic Performance of Uncertain Time-Varying Linear Systems},
  author={Evangelisti, L. L. and Pfifer, H.},
  booktitle={American Control Conf.},
  pages={1853--1858},
  year={2022}
}

@article{Satellite_2013,
  title={Satellite collision probability estimation using polynomial chaos expansions},
  author={Jones, B. A. and Doostan, A.},
  journal={Advances in Space Research},
  volume={52},
  number={11},
  pages={1860--1875},
  year={2013}
}

@article{Nagy_2007,
  title={Distributional uncertainty analysis using power series and polynomial chaos expansions},
  author={Nagy, Z. K. and Braatz, R. D.},
  journal={Journal of Process Control},
  volume={17},
  number={3},
  pages={229--240},
  year={2007}
}

@article{parametric_problems_2020,
  title={Polynomial chaos expansion for parametric problems in engineering systems: A review},
  author={Shen, D. and Wu, H. and Xia, B. and Gan, D.},
  journal={IEEE Systems Journal},
  volume={14},
  number={3},
  pages={4500--4514},
  year={2020}
}

@article{control_law_approximation_2019,
  title={Explicit {MPC} based on the {Galerkin} method for {AGC} considering volatile generations},
  author={Qiu, Y. and Lin, J. and Liu, F. and Song, Y.},
  journal={IEEE Trans. on Power Systems},
  volume={35},
  number={1},
  pages={462--473},
  year={2019}
}

@article{roa_2020,
  title={Region of attraction analysis of nonlinear stochastic systems using Polynomial Chaos Expansion},
  author={Ahbe, E. and Iannelli, A. and Smith, R. S.},
  journal={Automatica},
  volume={122},
  pages={109187},
  year={2020}
}

@article{moment_estimation_2020,
  title={On moment estimation from polynomial chaos expansion models},
  author={Lefebvre, T.},
  journal={IEEE Control Systems Letters},
  volume={5},
  number={5},
  pages={1519--1524},
  year={2020}
}

@article{Data_driven_stability_TAC_2020,
  title={Data-driven model predictive control with stability and robustness guarantees},
  author={Berberich, J. and K{\"o}hler, J. and M{\"u}ller, M. A and Allg{\"o}wer, F.},
  journal={IEEE Trans. on Autom. Control},
  volume={66},
  number={4},
  pages={1702--1717},
  year={2020}
}

@inproceedings{Lygeros_ECC_2019,
  title={Data-enabled predictive control: In the shallows of the {DeePC}},
  author={Coulson, J. and Lygeros, J. and D{\"o}rfler, F.},
  booktitle={European Control Conf.},
  pages={307--312},
  year={2019}
}

@article{Linear_behavioral_2022,
  title={Linear tracking {MPC} for nonlinear systems—Part I: The model-based case},
  author={Berberich, J. and K{\"o}hler, J. and M{\"u}ller, M. A. and Allg{\"o}wer, F.},
  journal={IEEE Trans. on Autom. Control},
  volume={67},
  number={9},
  pages={4390--4405},
  year={2022}
}

@article{Koopman_behavioral_Colin,
  title={Koopman based data-driven predictive control},
  author={Lian, Y. and Wang, R. and Jones, C. N.},
  journal={arXiv preprint arXiv:2102.05122},
  year={2021}
}

@article{fundamental_lemma_2005,
  title={A note on persistency of excitation},
  author={Willems, J. C. and Rapisarda, P. and Markovsky, I. and De Moor, B. L. M.},
  journal={Systems \& Control Letters},
  volume={54},
  number={4},
  pages={325--329},
  year={2005}
}

@book{Fisher_2008,
  title={Stability analysis and control of stochastic dynamic systems using polynomial chaos},
  author={Fisher, J. R.},
  year={2008},
  publisher={Texas A\&M University, USA}
}

@article{RPI_2005,
  title={Invariant approximations of the minimal robust positively invariant set},
  author={Rakovic, S. V. and Kerrigan, E. C. and Kouramas, K. I. and Mayne, D. Q.},
  journal={IEEE Trans. on Autom. Control},
  volume={50},
  number={3},
  pages={406--410},
  year={2005}
}

@inproceedings{Farina_CDC_2013,
  title={A probabilistic approach to model predictive control},
  author={Farina, M. and Giulioni, L. and Magni, L. and Scattolini, R.},
  booktitle={IEEE Conf. on Decision and Control},
  pages={7734--7739},
  year={2013}
}

@inproceedings{Imsland_2019,
  title={Output feedback stochastic nonlinear model predictive control of a polymerization batch process},
  author={Bradford, E. and Reble, M. and Imsland, L.},
  booktitle={European Control Conf.},
  pages={3144--3151},
  year={2019}
}

@inproceedings{pseudo_spectral_collocation_22,
  title={Stochastic Model Predictive Control of Discrete-time Cooperative Unmanned System},
  author={Bhusal, R. and Bhattacharjee, D. and Subbarao, K.},
  booktitle={AIAA Scitech Forum},
  pages={0961},
  year={2022}
}

@inproceedings{Rolf_CDC_2015,
  title={Efficient stochastic model predictive control based on polynomial chaos expansions for embedded applications},
  author={Lucia, S. and Zometa, P. and K{\"o}gel, M. and Findeisen, R.},
  booktitle={IEEE Conf. on Decision and Control},
  pages={3006--3012},
  year={2015}
}

@article{Faulwasser_TAC_2022,
  title={On a stochastic fundamental lemma and its use for data-driven optimal control},
  author={Pan, G. and Ou, R. and Faulwasser, T.},
  journal={IEEE Trans. on Autom. Control},
  year={2022}
}

@inproceedings{von_ACC_2020,
  title={Stochastic dynamic optimization and model predictive control based on polynomial chaos theory and symbolic arithmetic},
  author={von Andrian, M. and Braatz, R. D.},
  booktitle={American Control Conf.},
  pages={3399--3404},
  year={2020}
}

@article{Imsland_2021,
  title={Combining {Gaussian} processes and polynomial chaos expansions for stochastic nonlinear model predictive control},
  author={Bradford, E. and Imsland, L.},
  journal={arXiv preprint arXiv:2103.05441},
  year={2021}
}

@article{Stochastic_simulation_Khammas_2012,
  title={Simulation of stochastic systems via polynomial chaos expansions and convex optimization},
  author={Fagiano, L. and Khammash, M.},
  journal={Physical Review E},
  volume={86},
  number={3},
  pages={036702},
  year={2012}
}

@article{haptic_20,
  title={Modulation of control authority in adaptive haptic shared control paradigms},
  author={Izadi, V. and Ghasemi, A. H.},
  journal={Mechatronics},
  volume={78},
  year={2021}
}

@article{bioreactor_22,
  title={Model development and optimal control of a continuous packed bed bioreactor for the production of succinic acid},
  author={Zacharopoulos, I. and Tao, M. and Theodoropoulos, C.},
  journal={IFAC-PapersOnLine},
  volume={55},
  number={20},
  pages={493--498},
  year={2022}
}

@inproceedings{Reservoir_22,
  title={The Factor of Safety-Constrained Model Predictive Controller Design for Closed-Loop Reservoir Management},
  author={Ganesh, A. and Chalaturnyk, R. and Prasad, V.},
  booktitle={International Symposium on Advanced Control of Industrial Processes},
  pages={24--29},
  year={2022}
}

@article{propulsion_system_22,
  title={A Polynomial-Chaos-Based Multifidelity Approach to the Efficient Uncertainty Quantification of Online Simulations of Automotive Propulsion Systems},
  author={Yang, H. and Gorodetsky, A. and Fujii, Y. and Wang, K.},
  journal={Journal of Computational and Nonlinear Dynamics},
  volume={17},
  number={5},
  year={2022}
}

@article{EV_battery_22,
  title={Stochastic model predictive control for optimal charging of electric vehicles battery packs},
  author={Pozzi, A and Raimondo, D. M.},
  journal={Journal of Energy Storage},
  volume={55},
  year={2022}
}

@article{hybrid_update_Allgower,
  title={Recursive Feasibility and Stability for Stochastic {MPC} based on Polynomial Chaos},
  author={Ma, Z. and Schl{\"u}ter, H. and Berkel, F. and Specker, T. and Allg{\"o}wer, F.},
  journal={IFAC-PapersOnLine},
  volume={56},
  number={1},
  pages={204--209},
  year={2023}
}

@article{Xiu_Karniadakis_PCE_tutorial,
  title={The {Wiener--Askey} polynomial chaos for stochastic differential equations},
  author={Xiu, D. and Karniadakis, G. E.},
  journal={SIAM Journal on Scientific Computing},
  volume={24},
  number={2},
  pages={619--644},
  year={2002}
}

@book{numerical_analysis_2003,
  title={An Introduction to Numerical Analysis},
  author={S{\"u}li, E. and Mayers, D. F.},
  year={2003},
  publisher={Cambridge University Press}
}

@book{Spectral_UQ_2010,
  title={Spectral Methods for Uncertainty Quantification: with Applications to Computational Fluid Dynamics},
  author={Le Ma{\^\i}tre, O. and Knio, O. M.},
  year={2010},
  publisher={Springer Science \& Business Media}
}

@inproceedings{Lucia_ACC_2017,
  title={On stability of stochastic linear systems via polynomial chaos expansions},
  author={Lucia, S. and Paulson, J. A. and Findeisen, R. and Braatz, R. D.},
  booktitle={American Control Conf.},
  pages={5089--5094},
  year={2017}
}

@inproceedings{fagiano_CDC_2012,
  title={Nonlinear stochastic model predictive control via regularized polynomial chaos expansions},
  author={Fagiano, L. and Khammash, M.},
  booktitle={IEEE Conf. on Decision and Control},
  pages={142--147},
  year={2012}
}

@article{PDQ-LCSS, 
	author={Mishra, P. K. and Chatterjee, D. and Quevedo, D. E. }, 
	title={Output feedback stable stochastic predictive control with hard control constraints}, 
	journal={IEEE Control Systems Letters},
	year={2017}, 
	Volume = {1},
	Issue = {2},
	pages = {382 - 387}
}

@article{Wan2023,
   author = {Y. Wan and D. E. Shen and S. Lucia and R. Findeisen and R. D. Braatz},
   issn = {15582523},
   issue = {1},
   journal = {IEEE Trans. on Autom. Control},
   month = {1},
   pages = {470-477},
   title = {A Polynomial Chaos Approach to Robust $H_\infty$ Static Output-Feedback Control with Bounded Truncation Error},
   volume = {68},
   year = {2023}
}

@article{Wan2021,   
   author = {Y. Wan and D. E. Shen and S. Lucia and R. Findeisen and R. D. Braatz},   
   journal = {Automatica},
   title = {Polynomial Chaos-Based $H_2$ Output-Feedback Control of Systems with Probabilistic Parametric Uncertainties},
   volume = {131},
   year = {2021}
}

@inproceedings{wan_CDC_2017,
  title={A piecewise polynomial chaos approach to stochastic linear quadratic regulation for systems with probabilistic parametric uncertainties},
  author={Wan, Y. and Harinath, E. and Braatz, R. D.},
  booktitle={IEEE Conf. on Decision and Control},
  pages={505--510},
  year={2017}
}

@article{OHagan_SIAM_2013,
  title={Polynomial chaos: A tutorial and critique from a statistician’s perspective},
  author={O’Hagan, A.},
  journal={SIAM/ASA J. Uncertainty Quantification},
  volume={20},
  pages={1--20},
  year={2013}
}

@article{cameron1947,
  title={The orthogonal development of non-linear functionals in series of {Fourier-Hermite} functionals},
  author={Cameron, R. H. and Martin, W. T.},
  journal={Annals of Mathematics},
  pages={385--392},
  year={1947}
}

@article{Newman_Neurips_2021, 
   author = {B. Lütjens and C. H. Crawford and M. Veillette and D. Newman},
   journal = {Conf. on Neural Information Processing Systems},
   title = {Spectral PINNs: Fast Uncertainty Propagation with Physics-Informed Neural Networks},
   year = {2021}
}

@article{Kim_magazine_2013, 
   author = {K. K. Kim and D. E. Shen and Z. K. Nagy and R. D. Braatz},   
   journal = {IEEE Control Systems Magazine},
   pages = {58-67},
   title = {Wiener's Polynomial Chaos for the Analysis and Control of Nonlinear Dynamical Systems with Probabilistic Uncertainties [Historical Perspectives]},
   volume = {33},
   year = {2013},
}

@inproceedings{prandini2012randomized,
  title={A randomized approach to stochastic model predictive control},
  author={Prandini, M. and Garatti, S. and Lygeros, J.},
  booktitle={IEEE Conf. on Decision and Control},
  pages={7315--7320},
  year={2012},
  organization={IEEE}
}

% \endarticle

\end{document}